\documentclass[twocolumn]{aastex63}

\usepackage{amsmath,amssymb,amstext}
\usepackage{gensymb}
\usepackage{multirow}
\usepackage{rotating}
\usepackage{graphicx}

\usepackage{colortbl}

\graphicspath{{./}{Figures/}}

\newcommand{\spitzer}{{\textit{Spitzer}}}

\shorttitle{Spitzer Phase Curves}
\shortauthors{Davenport et al.}

\begin{document}

\title{An Analysis of Spitzer Phase Curves for WASP-121b and WASP-77Ab}

\correspondingauthor{Brian Davenport}
\email{bdav@umd.edu}

\author[0009-0000-7367-5541]{Brian Davenport}
\affiliation{Department of Astronomy, University of Maryland, College Park, MD 20742, USA}

\author[0000-0002-2984-3250]{Thomas Kennedy}
\affiliation{Department of Astronomy, University of Michigan, Ann Arbor, MI 48109, USA}

\author[0000-0002-2739-1465]{E. M. May}
\affiliation{Johns Hopkins APL, Laurel, MD 20723, USA}

\author[0000-0003-3963-9672]{Emily Rauscher}
\affiliation{Department of Astronomy, University of Michigan, Ann Arbor, MI 48109, USA}

\author[0000-0002-1337-9051]{Eliza M.-R. Kempton}
\affiliation{Department of Astronomy, University of Maryland, College Park, MD 20742, USA}

\author[0000-0003-0217-3880]{Isaac Malsky}
\affiliation{Department of Astronomy, University of Michigan, Ann Arbor, MI 48109, USA}

\author[0000-0002-7352-7941]{Kevin B. Stevenson}
\affiliation{Johns Hopkins APL, Laurel, MD 20723, USA}

\author[0000-0003-4733-6532]{Jacob L. Bean}
\affiliation{Department of Astronomy \& Astrophysics, University of Chicago, Chicago, IL 60637, USA}

\author[0000-0003-4241-7413]{Megan Weiner Mansfield}
\altaffiliation{NHFP Sagan Postdoctoral Fellow}
\affiliation{Steward Observatory, University of Arizona, Tucson, AZ 85719, USA}

\begin{abstract}
We present analyses of Spitzer InfraRed Array Camera (IRAC) 3.6 $\micron$ and 4.5 µm phase curve observations of hot Jupiters WASP-77Ab and WASP-121b. For WASP-121b, we find amplitudes of 1771 $\pm$ 95 ppm (3.6 µm) and 2048 $\pm$ 109 ppm (4.5 µm), and near-zero  offsets of -0.78 $\pm$ 1.87$\degree$ (3.6 µm) and 0.42 $\pm$ 1.74$\degree$ (4.5 µm), consistent within 2.2$\sigma$ and 1.3$\sigma$, respectively, with JWST NIRSpec results. For WASP-77Ab, we find  amplitudes of 535 $\pm$ 52 ppm (3.6 µm) and 919 $\pm$ 40 ppm (4.5 µm), and offsets of 33.45 $\pm$ 2.79$\degree$ (3.6 µm) and 16.28 $\pm$ 2.52$\degree$ (4.5 µm). We report day- and nightside brightness temperatures: for WASP-121b, 2779 $\pm$ 40 K (3.6 µm) and 2905 $\pm$ 51 K (4.5 µm) (day) and 1259 $\pm$ 67 K (3.6 µm) and 1349 $\pm$ 54 K (4.5 µm) (night), and for WASP-77Ab, 1876 $\pm$ 23 K (3.6 µm) and 1780 $\pm$ 25 K (day) and 1501 $\pm$ 22 K (3.6 µm) and 1234 $\pm$ 20 K (4.5 µm) (night). Comparing WASP-121b data to general circulation models, we find evidence for drag inhibiting day-to-night heat transfer, which our model reproduces using magnetic circulation. Further, comparing both planets' data to Energy Balance Models, we show suppressed circulation in WASP-121b and potential evidence for an unusually high Bond albedo in WASP-77Ab. We add both planets to the Spitzer population study of previously identified trends in offset versus orbital period, finding that a positive trend is weakened, but not eliminated, by including these planets. 
\\
\end{abstract}

\section{Introduction}

Observing exoplanet phase curves remains a key method for characterizing atmospheres, especially for hot Jupiters, where tidal synchronization and poor heat redistribution contribute to large differences in dayside and nightside emission \citep{Cowan2011}, and short orbital periods allow for observations of full orbits. Wavelength-dependent parameters derived from phase curves, such as secondary eclipse depth, phase amplitude, and phase offset, give insight into atmospheric dynamics. Phase Curves have been utilized to help constrain the presence and magnitude of such effects as equatorial wind jets \citep{Tan2020}, atmospheric drag \citep{May2021}, and cloud formation \citep{{Keating2019}, {Beatty2019}}, primarily through comparisons to simulated curves produced by General Circulation Models (GCM) run both with and without these effects present.

With Spitzer's warm mission complete, several studies of population trends from sets of its phase curves have looked at how phase amplitudes and offsets vary with planetary parameters such as period \citep{May2022}, and equilibrium temperature and stellar effective temperature \citep{Bell2021}. These studies, however, have yet to leverage the full set of observed Spitzer phase curves. For instance, the tentative empirical trends of increased offset with orbital period identified by \cite{May2022} (from 14 planets) and decreased amplitude with stellar effective temperature suggested by \cite{Bell2021} (from 15 planets), both of which would warrant further theoretical study, could be more robustly evaluated with a larger sample size. 

One key advantage of observing phase curves over secondary eclipses is the ability to probe nightside thermal emission. By observing both the flux during the secondary eclipse, which gives stellar flux without the planet, and the total flux near the primary transit, we can isolate the nightside flux contribution from the planet alone. In the future, this may also be extended to less time-intensive sparsely sampled phase curves by detecting planetary infrared excess from the nightside of exoplanets with far fewer data \citep{Stevenson2020}.  Nightside thermal emission also displays unique trends in Spitzer phase curve data -- in particular \cite{Beatty2019} and \cite{Keating2019} found that nightside emission departs from trends found in dayside emission, potentially due to the ubiquity of nightside cloud formation resulting in near uniform nightside temperatures across a range of planetary parameters. 

Recent phase curve observations by JWST have provided an opportunity to validate the phase curve observables derived from Spitzer data. Currently, JWST has completed phase curve observations for only four hot Jupiters (\cite{Mikal2023}; ERS 1336, PI Batalha; GO 2158, PI Parmentier; GO 2488, PI Sikora), requiring an average of over 31 hours of observing time each. Showing that phase curve observables are consistent between JWST and Spitzer will allow researchers to leverage the far larger Spitzer library for inclusion in population studies, which will be difficult with JWST alone.  \cite{Mikal2023} presented the first published hot Jupiter phase curve measurement from JWST, of WASP-121b. This Near Infrared Spectrograph (NIRSpec) G395H phase curve is split into two detectors (NRS1 and NRS2), which roughly align with the Spitzer channel 1 (3.6 $\micron$) and channel 2 (4.5 $\micron$) bandpasses. The phase offsets observed in NRS1 and NRS2 presented by \cite{Mikal2023} disagree by $>$3$\sigma$ with the channel 1 and channel 2 phase offsets for presented by \cite{Morello2023}, noting, however, that NRS1 and NRS2 do not perfectly overlap with the Spitzer bandpasses. Secondary eclipse depths, however, have so far shown better agreement between Spitzer and JWST. \cite{August2023} found eclipse depths for WASP-77Ab agreed within 2$\sigma$ between NRS1 and NRS2, and Spitzer channels 1 and 2. Because the greater length of observations for phase curves in comparison to eclipses expands the effects of instrument systematics, the increased difference in observables for the former may be dependent on reduction technique. Cross-comparisons of results across different reduction methods for Spitzer phase curves have shown significant variation in some cases \citep{Bell2021}, warranting the consideration of multiple systematic models when comparing across observatories.

In this paper, we present the first full phase curve analysis for WASP-77Ab as well as a new analysis of the Spitzer WASP-121b phase curve observations. As previous studies have shown the importance of using uniform data reduction techniques when analyzing population trends with Spitzer due to its significant systematics \citep{Bell2021, May2022}, we use Bilinearly-Interpolated Subpixel Sensitivity (BLISS) mapping of intrapixel sensitivity for uniformity to expand on the population trend analyses by \cite{May2022}. These two planets were selected due to their close orbital periods, for which we expect from the previously identified trend in orbital period to find similar phase offsets. Their disparate sizes, masses, and equilibrium temperatures help to test whether the observed trend, if supported, is truly independent of other key planetary parameters. 

As a part of this work we also explore the generation of a 3.6 $\micron$ fixed sensitivity map, following the methods in \cite{May2020}, noting that the population trends identified above were by necessity derived solely from 4.5 µm observations due to the high variability in  3.6 µm observables across reductions \citep{May2022}. Results of this are presented in Section \ref{sec:3.6micronmap}. In Section \ref{sec:methods}, we discuss the data reduction process and present best-fit parameter results.  In Section \ref{sec:GCMs}, we introduce the GCM methods employed and the results of simulations conducted with variations in the presence of clouds and of magnetic fields for both planets. We then compare the simulation results to our observations in Section \ref{sec:ModelDataComp}, and provide our conclusions in Section \ref{sec:Conclusions}. 

\section{Methods} \label{sec:methods}
\subsection{3.6 $\mu$m Fixed Sensitivity Map} \label{sec:3.6micronmap}
\cite{May2020} report issues with developing a fixed BLISS map for the 3.6 $\mu$m Spitzer channel. The authors found a time-dependence in intrapixel sensitivity using calibration data for the star KF09T1, which precluded development of a fixed map. We further investigated the calibration data to determine whether this time-dependence is a feature of the observed calibration star or of the detector channel itself.

\begin{figure}
    \centering
    \includegraphics[width=0.5\textwidth]{ 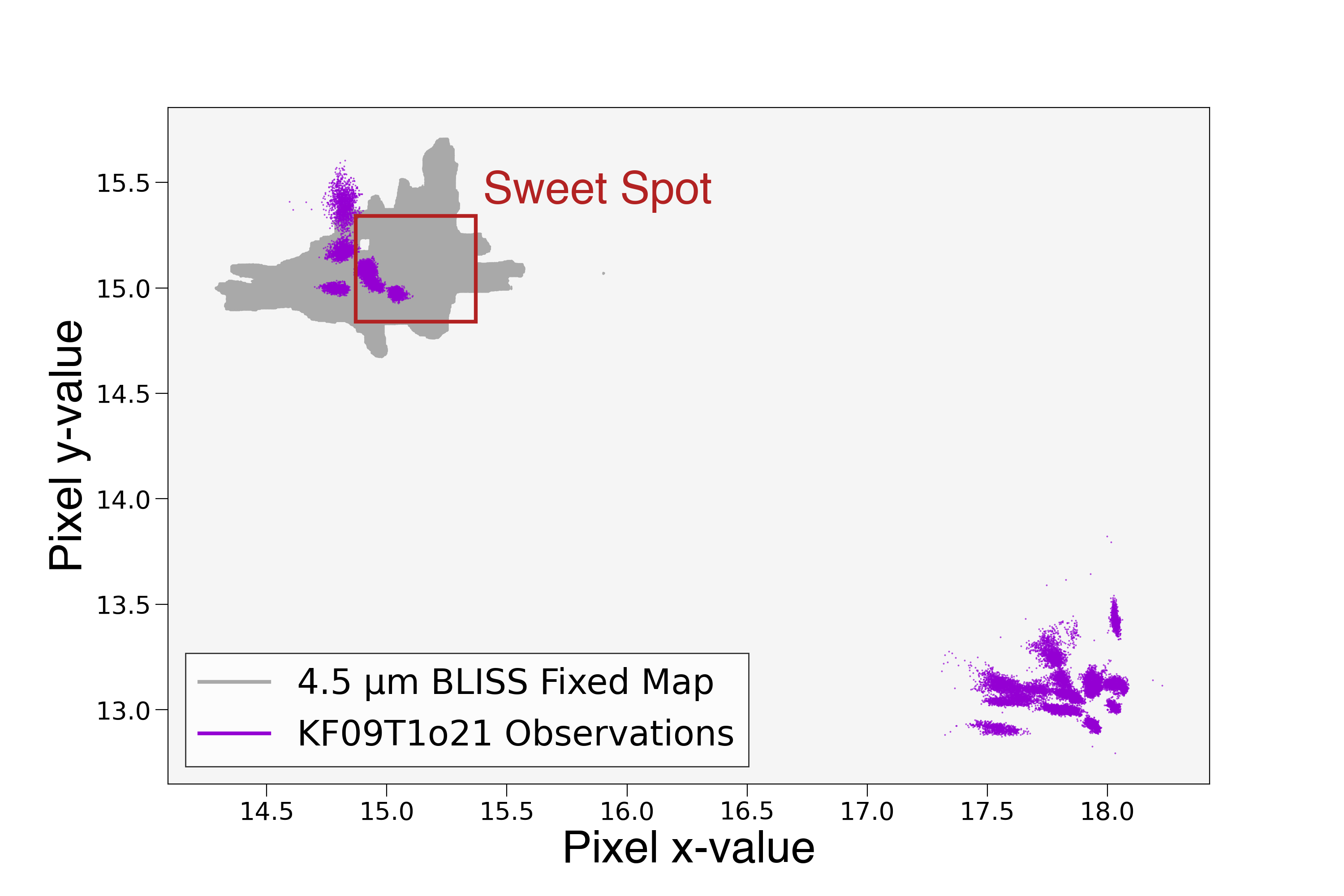}
    \caption{X and Y position of the 23 AORs observing KF09T1 with the 4.5 $\mu$m channel of Spitzer, showing that only a small fraction of the data overlapped with the fixed sensitivity map. The `Sweet Spot' marked on the plot is from \cite{Ingalls2012}.}
    \label{fig:Calibration}
\end{figure}

To test this, we analyzed calibration observations of KF09T1 at 4.5 $\mu$m with the intent to use the 4.5 $\micron$ fixed sensitivity BLISS map developed by \cite{May2020} to verify if evidence of stellar variability exists. Figure \ref{fig:Calibration} shows that the calibration star was not correctly centered on the ``sweet spot'', as defined by \cite{Ingalls2012}, for 18 out of 23 AORs, precluding the use of the fixed 4.5 $\micron$ map for a significant portion of the data. 

\begin{figure}
    \centering
    \includegraphics[width=0.45\textwidth]{ 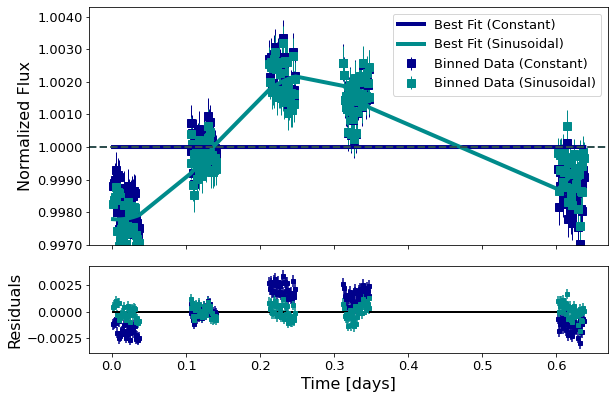}

    \caption{Comparison of constant (navy blue) versus sinusoidally varying (cyan) fits to the 4.5 $\mu$m observations of KF09T1 which overlapped with the 4.5 $\micron$ fixed sensitivity map. Evident in the plots is the need for some amount of variability to model the stellar flux.}
    \label{fig:Cal Fit}
\end{figure}

Using the Photometry for Orbits, Eclipses, and Transits (\texttt{POET}; \cite{Campo2011}; \cite{Stevenson2012}; \cite{Cubillos2013}) data reduction pipeline, we reduced the 5 AORs that overlap the 4.5 $\mu$m fixed map to compare the fit of a constant flux value with a sinusoidal varying model (as a proxy for stellar variability). Figure \ref{fig:Cal Fit} shows both of these fits compared to the KF09T1 calibration data after removing the intrapixel effect using the fixed 4.5 $\micron$ map. The sinusoidal model was strongly preferred by a Bayesian Information Criterion (BIC) comparison, providing an indication of some form of variability in the stellar flux; however, because of the limited data, this variability could not be assessed with confidence nor could it be well-quantified.

Since Spitzer is no longer operational, and no other calibration observations were taken with sufficient data to cover the full sweet spot at 3.6 $\mu$m, it is infeasible to develop a fixed sensitvity map for BLISS at 3.6 $\mu$m in the same manner as the \cite{May2020} 4.5 $\micron$ map. Therefore, all 3.6 $\micron$ data in this work are fit using a free BLISS map.

\subsection{Observation} \label{ssec:observations}

\begin{deluxetable*}{c c c c c c c l }
    \tablecolumns{8}
    \tabletypesize{\footnotesize}
    \caption{Planet Parameters}
    \label{table:planet params}
    \tablehead{
    \colhead{Planet} &
        \colhead{a} &
        \colhead{log(g)} &
        \colhead{Radius} &
        \colhead{Mass} &
        \colhead{T$_{eq}$} &
        \colhead{Period} &
        \colhead{Reference} \\
         {} & (AU) & (cm/s$^2$) & (R$_{\rm{Jup}}$) & (M$_{\rm{Jup}}$) & (K) & (days)  &  }
    \startdata
        \hline \hline
        WASP-121b   & 0.02596 $\pm$ 0.00063 & 2.970 $^{+0.033}_{-0.031}$ & 1.753 $\pm$ 0.036 & 1.157 $\pm$ 0.070 &  2503 $\pm$ 83 & 1.27492504 $\pm$ 1.5E-7  & \cite{Bourrier2020}   \\ \hline
        WASP-77Ab  & 0.02335 $\pm$ 0.00045 & 3.438 $^{+0.012}_{-0.016}$ & 1.230 $\pm$ 0.031 & 1.667 $^{+0.068}_{-0.064}$ & 1715 $^{+26}_{-25}$ & 1.36002854 $\pm$ 6.2E-7   & \cite{Cortes2020}   \\  \hline
    \enddata
    \tablecomments{\cite{Bourrier2020} does not report an equilibrium temperature. T$_{eq}$ for WASP-121b is calculated from values in this table and Table \ref{table:LimbDarkening}, assuming a Bond albedo of zero, for consistency.}
\end{deluxetable*}

\subsubsection{WASP-121b}
\begin{figure*}
    \centering
    \includegraphics[width=0.4\textwidth]{ 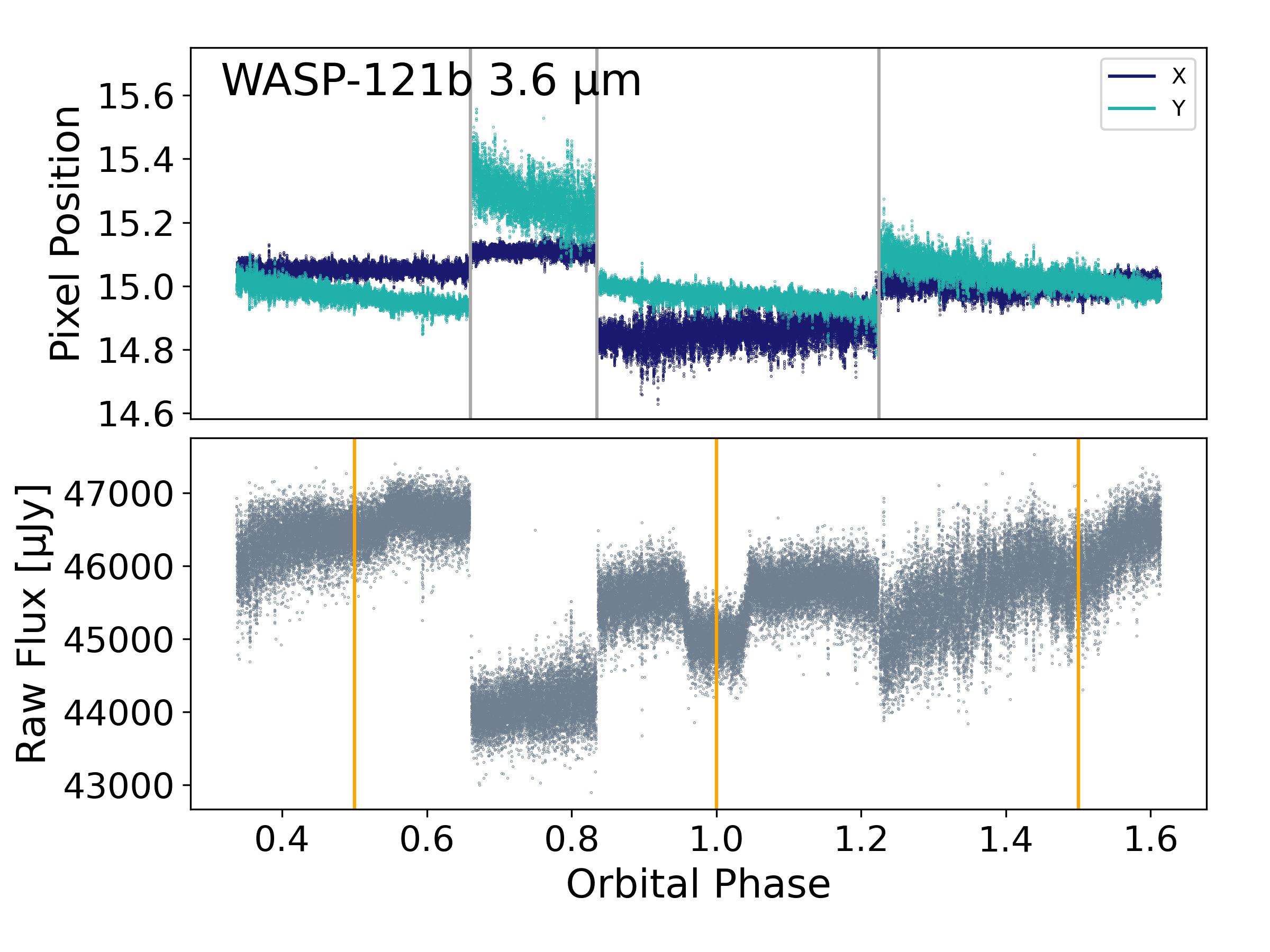}
    \includegraphics[width=0.4\textwidth]{ 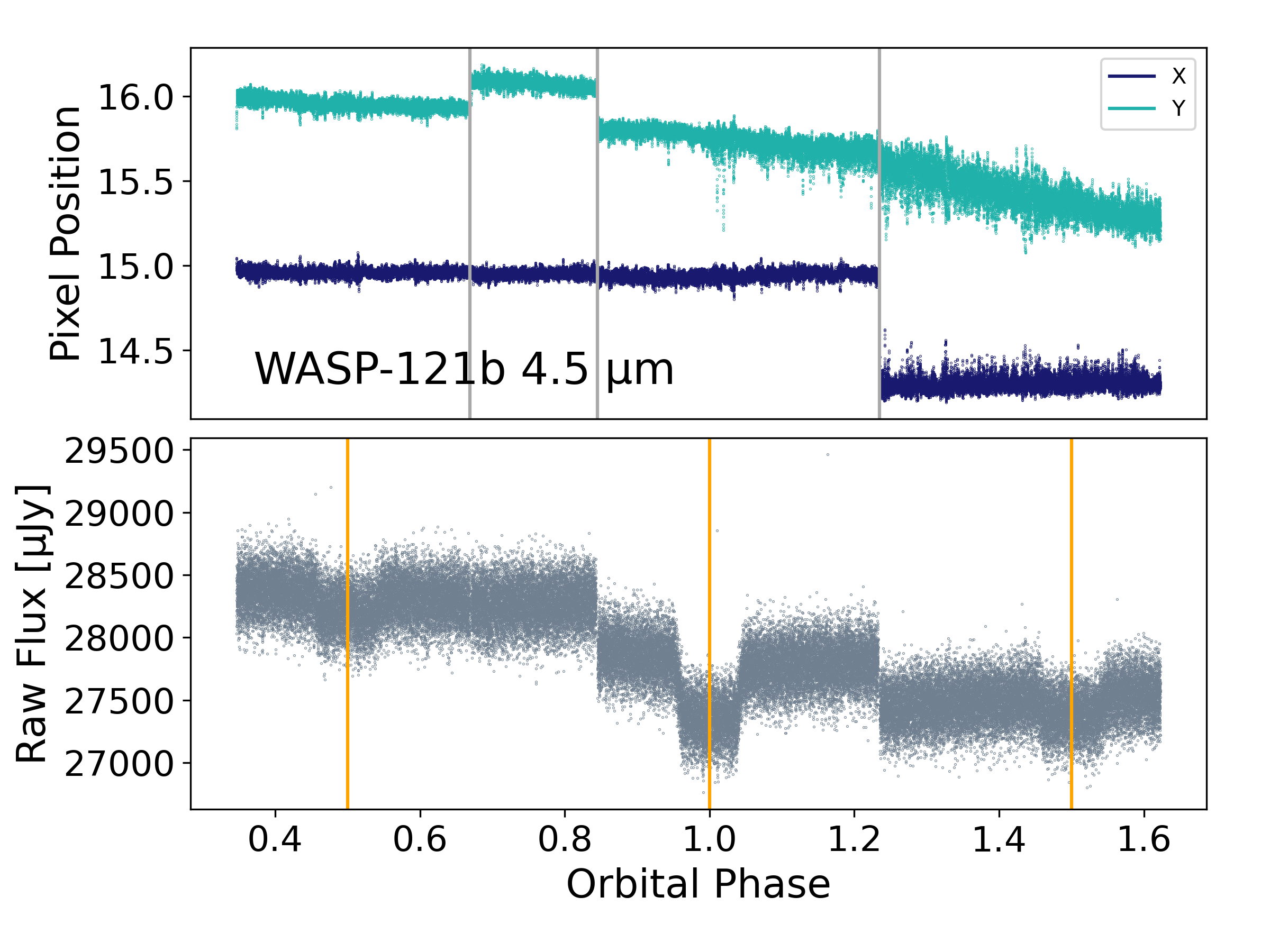}
    \includegraphics[width=0.4\textwidth]{ 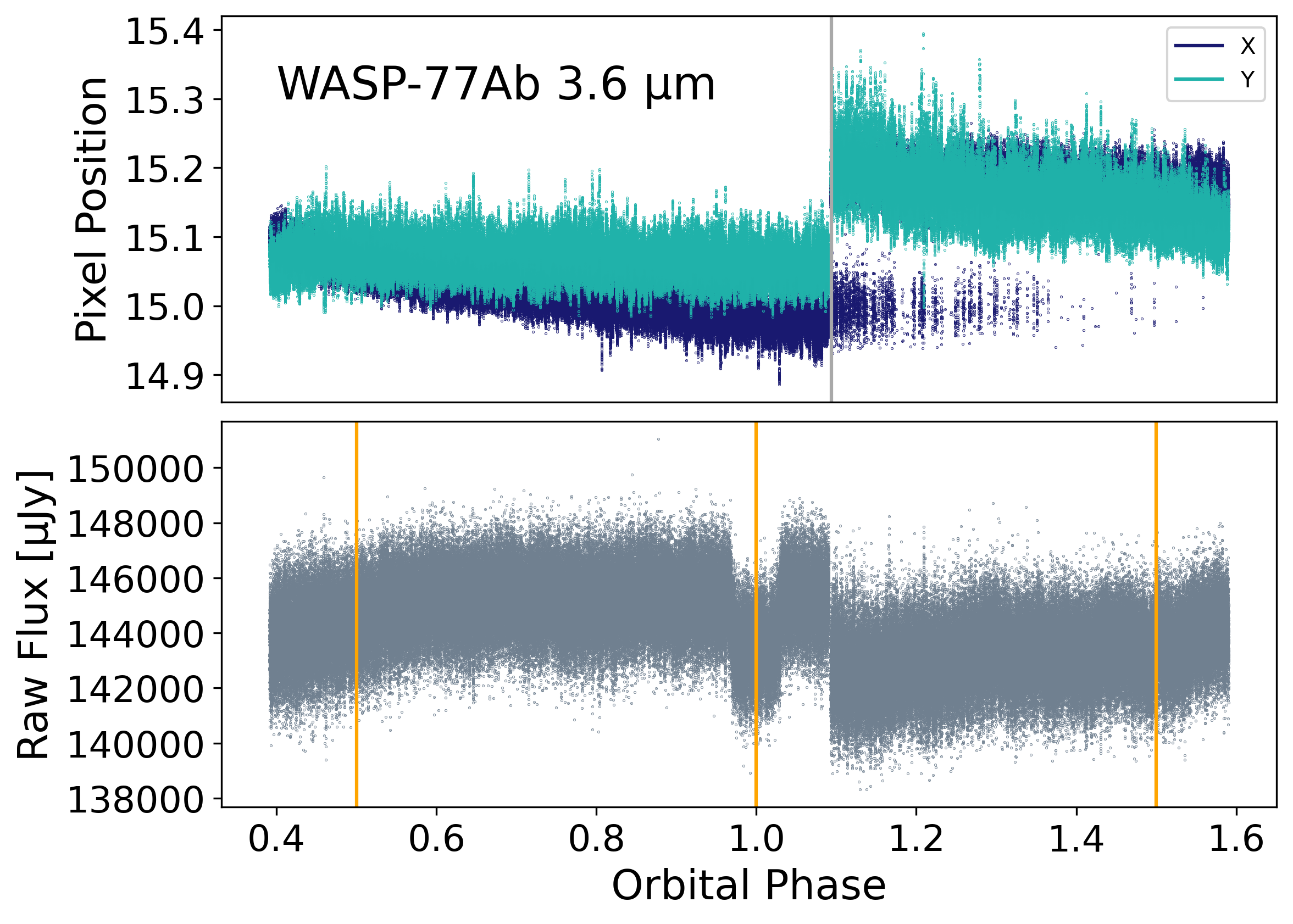}
    \includegraphics[width=0.4\textwidth]{ 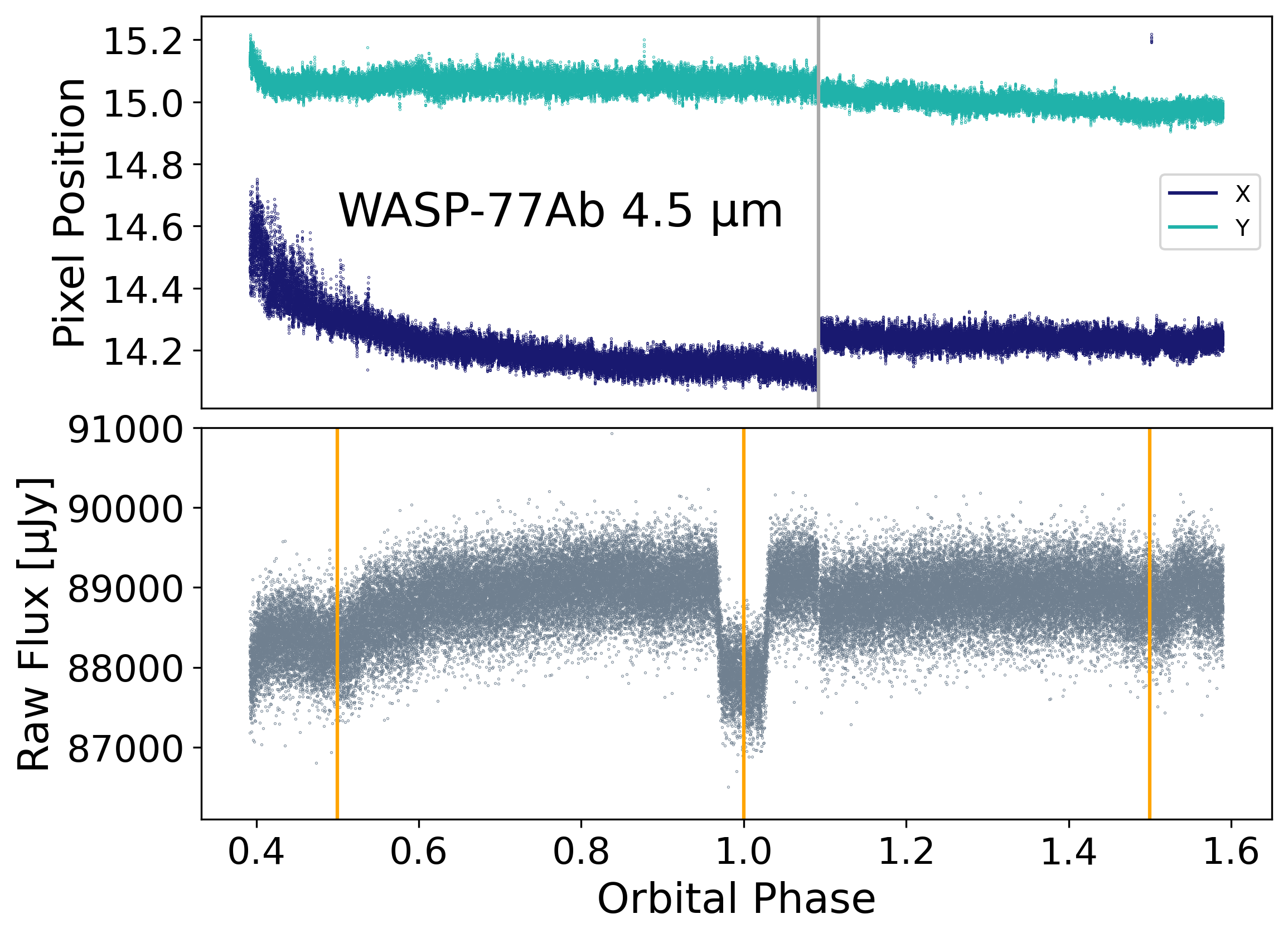}

    \caption{X and Y centroids (\textbf{top}) and corresponding raw (uncorrected) flux (\textbf{bottom}) for all phase curves analyzed in this study. In the top panel, the AOR gaps are marked by vertical lines and in the bottom panel phases of $\pm$ 0.5 (secondary eclipse) and 1.0 (transit) are marked to guide the eye. }
    \label{fig:Raw Flux}
\end{figure*}

WASP-121b is a $\sim$2500K, 1.2 M$_{\rm{Jup}}$ ultra-hot Jupiter orbiting an F-type star at a 1.27 day period. Additional details of the planet and star are in Tables \ref{table:planet params} and \ref{table:LimbDarkening}. Two phase curves were observed for WASP-121b using Spitzer’s InfraRed Array Camera (IRAC; \cite{Fazio2004}), one at channel 1 (3.6 $\mu$m), and one at channel 2 (4.5 $\mu$m) (Program 13242; PI: Tom Evans). Each phase curve covered approximately 40 hours with 2s frame times for both channels, as outlined in Table \ref{table:observations}. Each phase curve was separated into four astronomical observation requests (AORs). Figure \ref{fig:Raw Flux} shows the x- and y- centroids (top subpanel) and raw flux (bottom subpanel) for both WASP-121b phase curves. The planet has also been observed over full-orbit phase curves in the infrared by JWST NIRSpec \citep{Mikal2023} and the HST Wide Field Camera 3 (HST/WFC3) \citep{Mikal2022}, and in the optical by TESS \citep{Bourrier2020b}.

\begin{deluxetable*}{c c c c c c c c}
    \tablecolumns{6}
    \tabletypesize{\footnotesize}
    \caption{Stellar Parameters and Limb Darkening}
    \label{table:LimbDarkening}
    \tablehead{
    \colhead{Star} &
        \colhead{T$_{eff}$} &
        \colhead{log(\textit{g})} &
        \colhead{[Fe/H]} &
        \colhead{Radius} &
        \colhead{Ch.} &
        \colhead{C$_1$} &
        \colhead{C$_2$} \\
        & (K) & (cm/s$^2$) & (dex) & (R$_{\odot}$) & ($\micron$) & & }
    \startdata
        \hline \hline 
        WASP-121 & 6776 $\pm$ 138 & 4.234 $\pm$ 0.11 & 0.13 $\pm$ 0.09 & 0.910 $^{+0.025}_{-0.023}$ & 3.6   & 0.067  & 0.125 \\
                         &  &  &       &                   & 4.5   & 0.073  & 0.091 \\ \hline
        WASP-77A  & 5617 $\pm$ 72 & 4.476 $\pm$ 0.015 & -0.10 $\pm$ 0.11   &     1.458 $\pm$ 0.030  & 3.6 & 0.093 & 0.137 \\
                         &  &  &      &                    & 4.5   & 0.09  & 0.101 \\ \hline
    \enddata
    \tablecomments{C$_1$ and C$_2$ are the quadratic limb darkening parameters. WASP-121 temperature, gravity, metallicity, and radius are from \cite{Stassun2019}. WASP-77A values are from \cite{Cortes2020}. \\
     }
\end{deluxetable*}

\subsubsection{WASP-77Ab}
WASP-77Ab is a ~1700K, 2.0 M$_{\rm{Jup}}$ hot Jupiter orbiting a G-type star with a nearby K-type companion at a 1.36 day period. Additional details of the planet and star are in Tables \ref{table:planet params} and \ref{table:LimbDarkening}. One phase curve was observed each at 3.6 $\mu$m and 4.5 $\mu$m (Program 13038; PI: Kevin Stevenson) also for 40 hrs, with frame times of 0.4s at 3.6 $\mu$m, and 2.0s at 4.5 $\mu$m, resulting in approximately 5 times greater number of frames for WASP-77Ab channel 1 compared to the other three observations, as Table \ref{table:observations} shows. Each phase curve consisted of two AORs. Figure \ref{fig:Raw Flux} shows the x- and y- centroids (top subpanel) and raw flux (bottom subpanel) for both WASP-77Ab phase curves. This planet has not yet been observed by other instruments over a full phase, but has been observed in secondary eclipse by JWST NIRSpec \citep{August2023} and HST/WFC3 \citep{Mansfield2022}. WASP-77Ab has also been subject to atmospheric retrievals from both the JWST observation and from IGRINS thermal spectra, noteworthy for their confirmation of its subsolar metallicity \citep{{Line2021}, {Smith2024}}.

\begin{deluxetable*}{c c c  c c c}
    \tablecolumns{6}
    \tabletypesize{\footnotesize}
    \caption{Observations}
    \label{table:observations}
    \tablehead{
    \colhead{Label} &
        \colhead{Observation} &
        \colhead{Duration} \vspace{-0.2cm}&
        \colhead{Frame.} &
        \colhead{Total} &
        \colhead{Band} \\ 
          & \colhead{Date} & \colhead{(hrs)} & \colhead{Time (s)} & \colhead{Frames} & \colhead{($\micron$)}
                }
    \startdata
        \hline \hline
        wa121bo11   & Jan 29-30 2018  & 38.7    &  2.0  & 69,184    & 3.6   \\ \hline
        wa121bo21   & Jan 27-28 2018  & 38.7    &  2.0  & 69,184    & 4.5 \\ \hline
        wa77abo11   & Apr 10-11 2017  & 39.1    &  0.4  & 328,640    & 3.6   \\ \hline
        wa77abo21   & Nov 29-Dec 01 2016  & 39.0    &  2.0  & 69,504    & 4.5 \\  \hline
    \enddata
    \tablecomments{Label denotes the planet (e.g. wa121b = WASP-121b), type of observation (o=orbit), \spitzer\ IRAC channel (1 or 2) and visit number.}
\end{deluxetable*}

\subsection{Data Reduction}
We employed the \texttt{POET} data reduction pipeline for analysis. This method included 2D Gaussian centroiding and fixed (constant in time) aperture sizes selected for each observation by minimizing the standard deviation of the normalized residuals (SDNR) at 0.25 pixel increments for circular aperture sizes with radii between 2.0 and 4.0 pixels. The exception was the 4.5 $\mu$m channel observation for WASP-77Ab, which was optimized at 4.5 pixels due to the presence of companion star WASP-77B. The aperture size for the WASP-77Ab 3.6 $\mu$m observation was within the normal range but also large, at 3.5 pixels, for the same reason. The aperture size used for each observation is noted in Table \ref{table:bestfits}.

The 3.6 $\mu$m observation of WASP-77Ab did not sufficiently account for settling time, so a portion of the first AOR include data which are moving across the detector toward the ``sweet spot'' (a region of the pixel with roughly constant pixel sensitivity). This is clearly visible in Figure \ref{fig:BLISS}. To account for this, the data set was clipped by detector pixel position above 15.4 in the y direction and the first 8000 of the remaining frames, or just under an hour, were excluded for data analysis.

\subsection{Intrapixel Sensitivity} \label{ssec:intrapixel}

\subsubsection{WASP-121b BLISS Mapping}
\begin{figure*}
    \centering
    \includegraphics[width=0.4\textwidth]{ 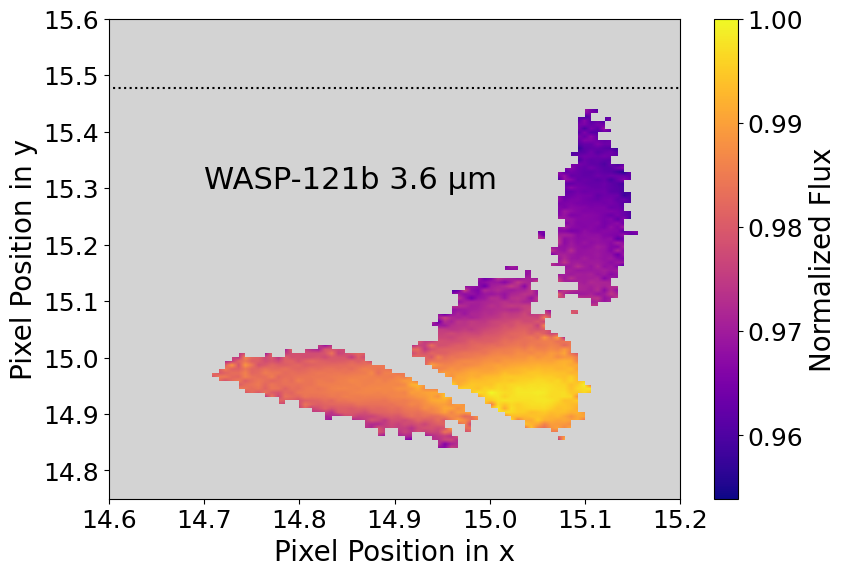}
    \includegraphics[width=0.4\textwidth]{ 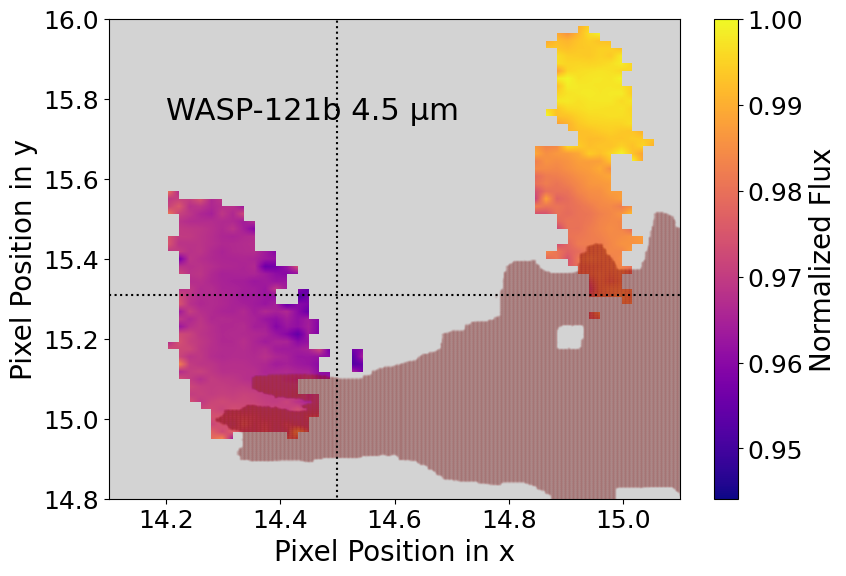}
    \includegraphics[width=0.4\textwidth]{ 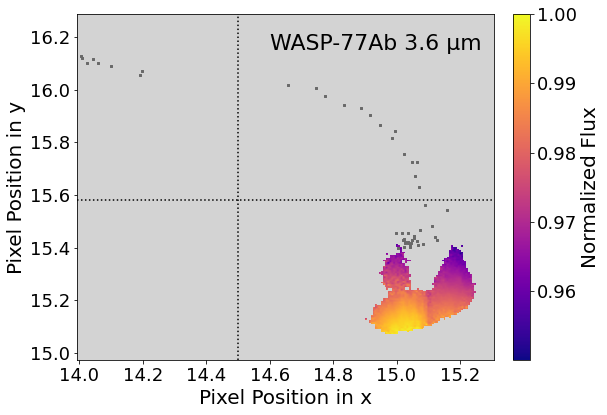}
    \includegraphics[width=0.4\textwidth]{ 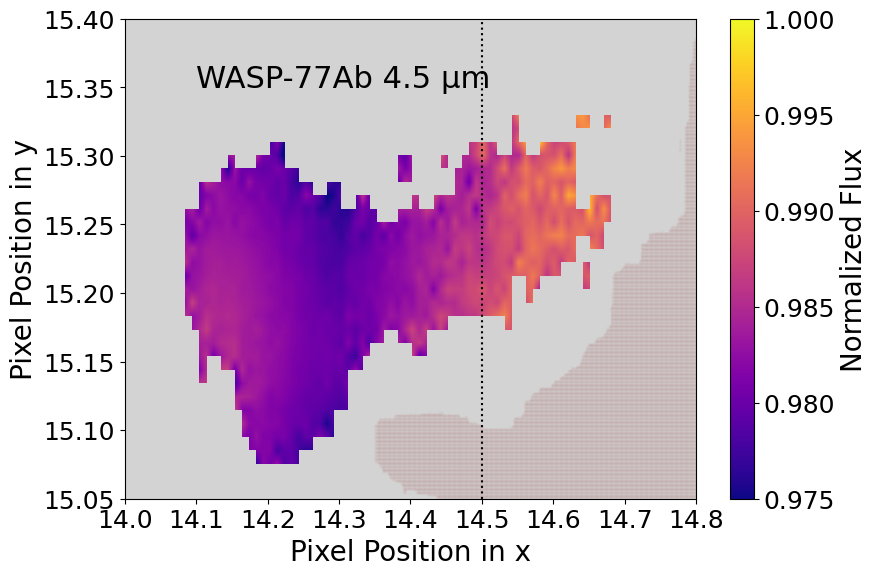}

    \caption{BLISS maps for the data sets analyzed in this work. The colorbar denotes the relative sensitivity of a given subpixel element. The 4.5 $\micron$ observations also show the extent of our fixed sensitivity map in the brown shaded region. The 3.6 $\mu$m map for WASP-77Ab (lower left) shows the pre-clipped data. The data removed for analysis are depicted in gray. Dashed lines denote the edges of a pixel (where relevant), with axis labels in sub-pixel units.}
    \label{fig:BLISS}
\end{figure*}

As shown in Figure \ref{fig:BLISS}, the 4.5 $\mu$m data for WASP-121b did not overlap with the fixed sensitivity map from \cite{May2020}, so a fixed map was not used. In addition to a free BLISS map, we employed Point Response Function (PRF) detrending to both channels to model the departure from circularity of the PRF as it moved from the center of a given pixel. We used BIC values to compare no PRF detrending with first, second, and third order polynomials of the Gaussian width in the x and y dimensions. Both channels preferred detrending, as denoted in Table \ref{table:bestfits}. Specifically the 3.6 $\micron$ phase curve is best fit by a 3$^{rd}$ order PRF function while the 4.5 $\micron$ phase curve is best fit by a 2$^{nd}$ order PRF function.
\subsubsection{WASP-77Ab BLISS Mapping}
For WASP-77Ab, the 4.5 pixel aperture size needed to obtain all the light from both WASP-77A and WASP-77B\footnote{A dilution correction to account for this is discussed in Section \ref{ssec:dilution}.} precluded the use of a fixed BLISS map, which only exists for the \texttt{POET} pipeline at apertures up to 4.0 pixels \citep{May2020}. Thus, we utilized free BLISS mapping for both channels. As with WASP-121b, BIC preferred PRF detrending to model the centroid, as denoted in Table \ref{table:bestfits}. The 3.6 $\micron$ phase curve is best fit by a 3$^{rd}$ order PRF function while the 4.5 $\micron$ phase curve is best fit by a 1$^{st}$ order PRF function.

\subsection{Astrophysical Source Models}
Quadratic limb darkening was incorporated into transit modeling using the \texttt{ExoCTK} limb-darkening tool \citep{exoctk} and Kurucz stellar models \citep{Kurucz2004}. Table \ref{table:LimbDarkening} shows the stellar parameters used for WASP-121 and WASP-77A and the calculated limb darkening parameters. To model the transits, we used the \texttt{BATMAN} package \citep{Kreidberg2015}, and for the secondary eclipses we use a model adapted from \cite{Mandel2002}. For the phase functions, all observations were fit with a sinusoidal variation in flux, considering both symmetric (full period sinusoids) and asymmetric (full period + half period sinusoids) phase functions. Table \ref{table:bestfits} gives an overview of the best fit model combinations for each phase curves, as further discussed in the following sections. To account for the potential presence of temporal ramps in the data, we tested various ramp models for each phase curve observation, using the minimized BIC to select the preferred model among a linear in time, exponential in time, or no temporal ramp. 

\subsubsection{WASP-121b}
The 3.6 $\mu$m data preferred an exponentially rising ramp, while the 4.5 $\mu$m data favored no temporal ramp. Comparative BIC values for different ramp models are displayed in Table \ref{table:bestfits}. In both cases a symmetric phase curve was preferred.

\subsubsection{WASP-77Ab}
For the 3.6 $\mu$m data of WASP-77Ab, only an exponential and an exponential coupled with linear ramp were compared as other ramp models produced nonphysical results (nightside flux ratios of less than 1). This was possibly due to the combined effect of instrument settling, discussed in section 2.2.2, with the detector ramp. Though the data were clipped to account for most of the drift in pointing, it has been noted that the ramp effect can persist longer due to charge trapping \citep{Lewis2013}. As with WASP-121b, the 4.5 $\mu$m data favored no temporal ramp. For the phase function, due to the appearance of a sharp rise in flux just prior to the eclipses for  3.6 $\mu$m, we also attempted to model those data with spherical harmonics using the \texttt{SPIDERMAN} package \citep{Louden2018}. The sinusoidal function, however, was still favored by BIC, and the comparisons are shown in Table \ref{table:bestfits}.

\begin{deluxetable}{c c c | l l l}
    \tablecolumns{6}
    \tabletypesize{\scriptsize}
    \caption{Best Fit Models}
    \label{table:bestfits}
    \tablehead{
    \colhead{Label} &
        \colhead{Aperture} & 
        \colhead{Systematic} \vspace{-0.2cm}&
        \colhead{Ramp} &
        \colhead{Phase} &
        \colhead{$\Delta$BIC} \\ 
         & \colhead{[Pixels]} & \colhead{Model} & \colhead{Model}  & \colhead{Model} &
        }
    \startdata
        \hline \hline
        wa121bo11   & 2.50  & Free     &  --       & Symm. & 366.21 \\
                    &   & BLISS         &  Lin.     & Symm. & 79.61  \\ 
                    &   &             &  \textbf{Exp.}    & \textbf{Symm.} & \textbf{0.0} \\
                    &   &  +              &  Exp. + Lin.     & Symm. & 9.12 \\ 
                    &   &      &  --       & Asymm. & 320.89 \\ 
                    &   &  3$^{rd}$        &  Lin.     & Asymm. & 115.91 \\ 
                    &   &  order           &  Exp.    & Asymm. & 76.17 \\
                    &   &  PRF             &  Exp. + Lin.     & Asymm. & 74.75 \\\hline \hline
        wa121bo21   & 2.50  & Free      &  \textbf{--}       & \textbf{Symm.} & \textbf{0.0} \\
                    &   & BLISS         &  Lin.     & Symm. & 10.33  \\ 
                    &   &  +            &  Exp.    & Symm. &  31.66\\ 
                    &   & 2$^{nd}$      &  --       & Asymm. & 13.02 \\ 
                    &   & order         &  Lin.     & Asymm. & 23.05 \\ 
                    &   & PRF           &  Exp.    & Asymm. & 44.22 \\ \hline
        wa77abo11   & 3.50  & Free BLISS & Exp. & Symm. & 63.36 \\    
                    & & + & Exp. + Lin.        & Symm. & 11.9  \\
                    &   &   3$^{rd}$ order           &  Exp.     & Asymm &  29.12 \\                  
                    &   &    PRF       &  \textbf{Exp. + Lin.}    & \textbf{Asymm.} &  \textbf{0}\\ 
                    \hline \hline 
        wa77abo21   & 4.50  & Free          &  \textbf{--}       & \textbf{Symm.} & \textbf{0.0} \\
                    &   & BLISS         &  Lin.     & Symm. & 10.55\\ 
                    &   & +             &  Exp.    & Symm.  & -- \\
                    &   & 1$^{st}$      &  --       & Asymm. & 18.47 \\ 
                    &   & order         &  Lin.     & Asymm. & 28.54 \\ 
                    &   & PRF           &   Exp.   & Asymm. & -- \\ \hline
    \enddata
    \tablecomments{Lin. = Linear Temporal Ramp; Exp. = Rising Exponential Temporal Ramp; Symm. = Symmetric sinusoidal phase function; Asymm. = Asymmetric sinusoidal phase function (additional half-period sinusoidal term).
   Some model combinations result in significantly nonphysical best fits (e.g., negative night side flux) and are not included. Bolded rows denote the best fit.}
\end{deluxetable}

\subsection{Dilution Correction} \label{ssec:dilution}
Because WASP-77A has a nearby stellar companion, WASP-77B, the measured flux for the WASP-77Ab observations included light from both stars. Based on WASP-77A parameters from Table \ref{table:LimbDarkening} and WASP-77B parameters from \cite{Maxted2013}, we used their blackbody fluxes to calculate dilution factors, $\alpha_{Comp}(\lambda)=F_{Comp}/F_{W77A}$, at each Spitzer channel as in \citep{Stevenson2014}. We obtained values of $0.414 \pm 0.013$ for 3.6 $\mu$m and $0.396 \pm 0.010$ for 4.5 $\mu$m and applied them to the relative flux measurements through the equation
\begin{equation} \label{dilution_factor}
\delta_{Corr}(\lambda) = [1 + g(\beta, \lambda)\alpha_{Comp}(\lambda)]\delta_{Meas}(\lambda)
\end{equation}
where $\delta_{Meas}(\lambda)$ and $\delta_{Corr}(\lambda)$ are measured and updated eclipse depths, respectively, and $g(\beta, \lambda)$ is a wavelength-dependent fraction of the companion flux within the observed aperture \citep{Stevenson2014}. Because of the large aperture sizes used (3.5 pixels at 3.6 $\mu$m and 4.5 pixels at 4.5 $\mu$m) and a pixel scale of $\sim$1.2 arcseconds/pixel compared to the ~3 arcsecond separation between the stars \citep{Maxted2013}, we approximate  $g(\beta, \lambda)$ as 1. We note that the choice of dilution factor primarily affects only the absolute values of hemisphere fluxes (see Table \ref{table:Results}). Phase offset is unaffected, while amplitude would change only by $\sim1\sigma$ without any dilution correction, and by much less with one calculated from different assumptions.

\begin{figure*}
    \centering
    \includegraphics[width=\textwidth]{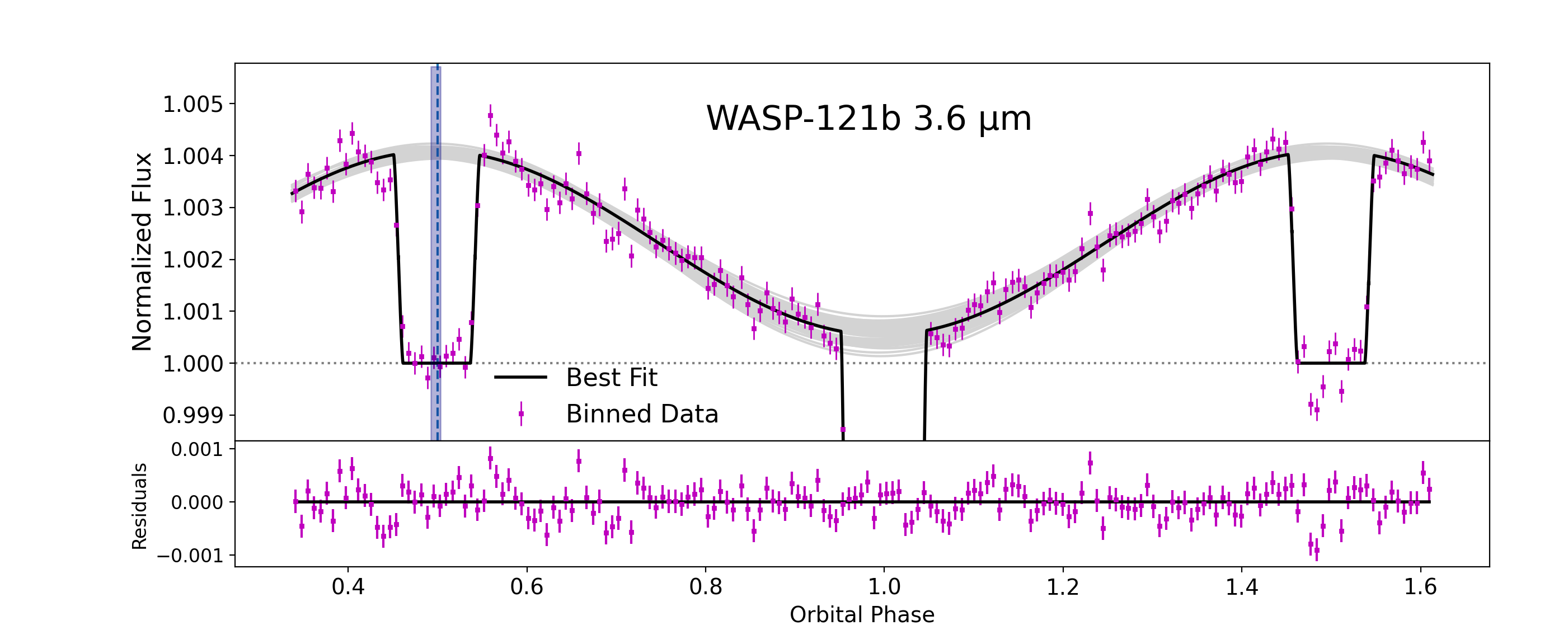}
    \includegraphics[width=\textwidth]{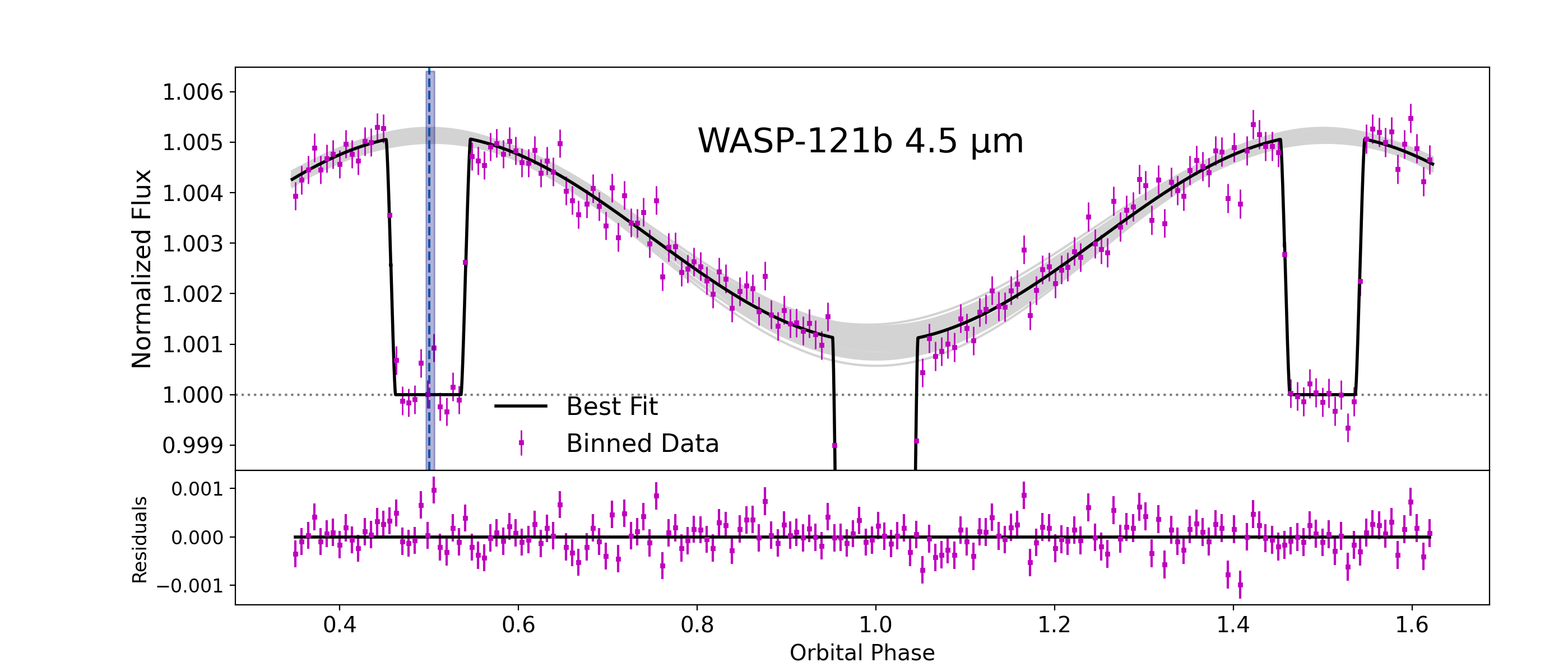}
    \caption{Best fit phase curves for WASP-121b at 3.6 $\mu$m (\textbf{top}) and 4.5 $\mu$m (\textbf{bottom}), with resulting residuals. The black curves show the best fit phase, eclipse and transit models. The gray curves are a sampling of 100 MCMC solutions for phase curve parameters to visualize the solution error. The dotted vertical line shows the center of first eclipse, and the vertical blue bar indicates calculated phase offset with error.}
    \label{fig:WASP-121b PC}
\end{figure*}

\subsection{Phase Curve Constraints} \label{ssec:phasecurves}
\subsubsection{WASP-121b}
With the 3.6 $\mu$m data we measured a phase offset of -0.78 $\pm$ 1.87$\degree$ and phase amplitude of 1771 $\pm$ 95 ppm. Though this does indicate a westward offset, it is consistent with 0 within 1$\sigma$, and agrees at 2.2$\sigma$ with the JWST NRS1 offset of 3.36 $\degree$ reported by \cite{Mikal2023}, compared to the previously reported 5.7 $\sigma$ discrepancy between \cite{Morello2023} and \cite{Mikal2023}. From a calculated eclipse depth of 4077 $\pm$ 59 ppm, we found a disk-integrated dayside brightness temperature of 2543 $\pm$ 42 K. Using the flux of the best fit phase curve at the center of primary transit, we found a disk-integrated nightside brightness temperature of 1082 $\pm$ 61 K. 

At 4.5 $\mu$m, we found a phase offset of 0.42 $\pm$ 1.74 $\degree$, which agrees to 1.3$\sigma$ with the \cite{Mikal2023} NRS2 phase offset of 2.66$\degree$, versus the previously reported 2.5$\sigma$ discrepancy between \cite{Morello2023} and \cite{Mikal2023}. Using a best-fit eclipse depth of 5121 $\pm$ 76 ppm, we calculated a dayside brightness temperature of 2905 $\pm$ 54 K at 4.5 µm. On the nightside we found a brightness temperature of 1349 $\pm$ 54 K. These values are summarized in Table \ref{table:Results}, and images of the fit of modeled flux to data for both channels are shown in Figure \ref{fig:WASP-121b PC}. 

\subsubsection{WASP-77Ab}
At 3.6 $\mu$m, we found a phase offset of 12.38$\pm$ 2.12$\degree$ and a dilution corrected phase amplitude of 1188$\pm$ 40 ppm. Although there exists no published JWST phase curve for this planet, \cite{August2023} calculated tentative phase offsets for NRS1 and NRS2 from an extended secondary eclipse observation. The 3.6 $\mu$m phase offset agrees within 0.6$\sigma$ with the NRS1 estimation. \cite{August2023} already showed agreement between Spitzer and JWST secondary eclipse depths within 2$\sigma$. From a dilution-corrected eclipse depth of 2417 $\pm$ 135 ppm, we calculated a dayside brightness temperature of 1403 $\pm$ 18 K. 

\begin{figure*}
    \centering
    \includegraphics[width=\textwidth]{ 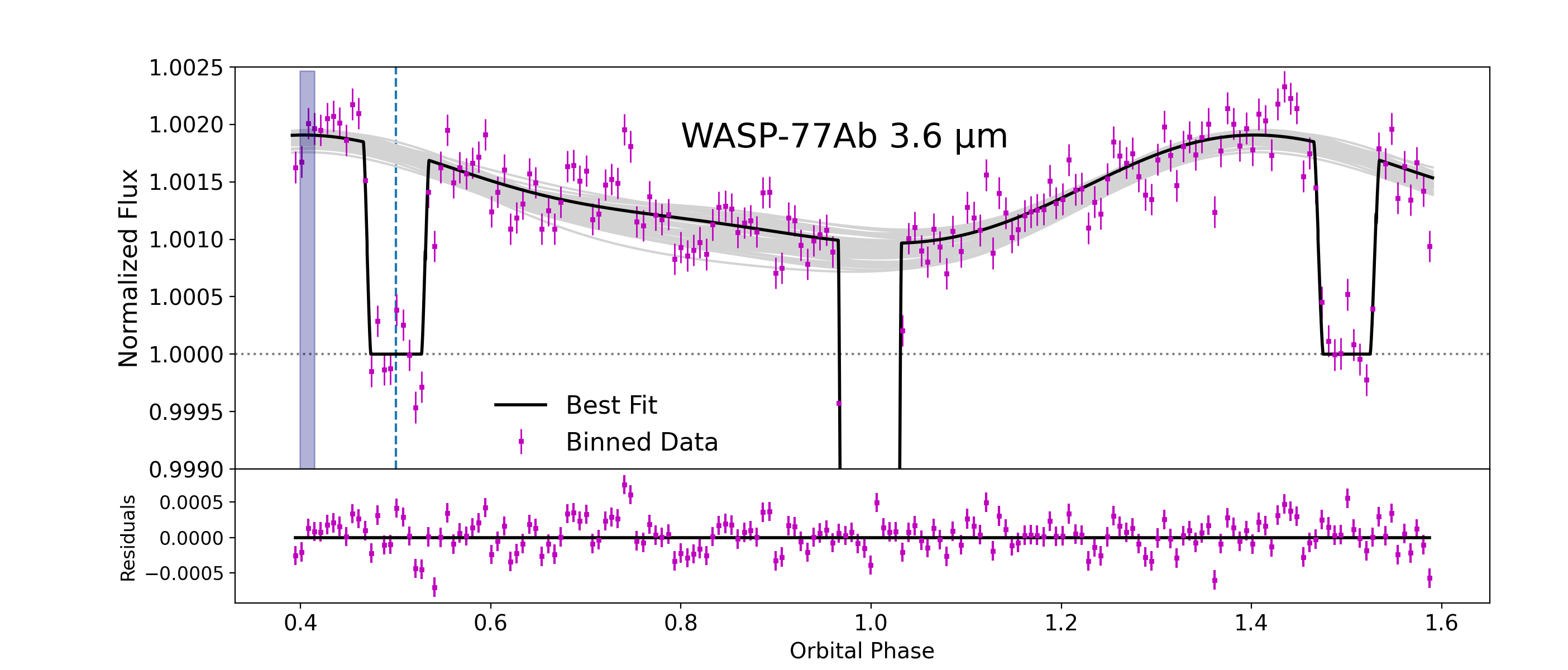}
    \includegraphics[width=\textwidth]{ 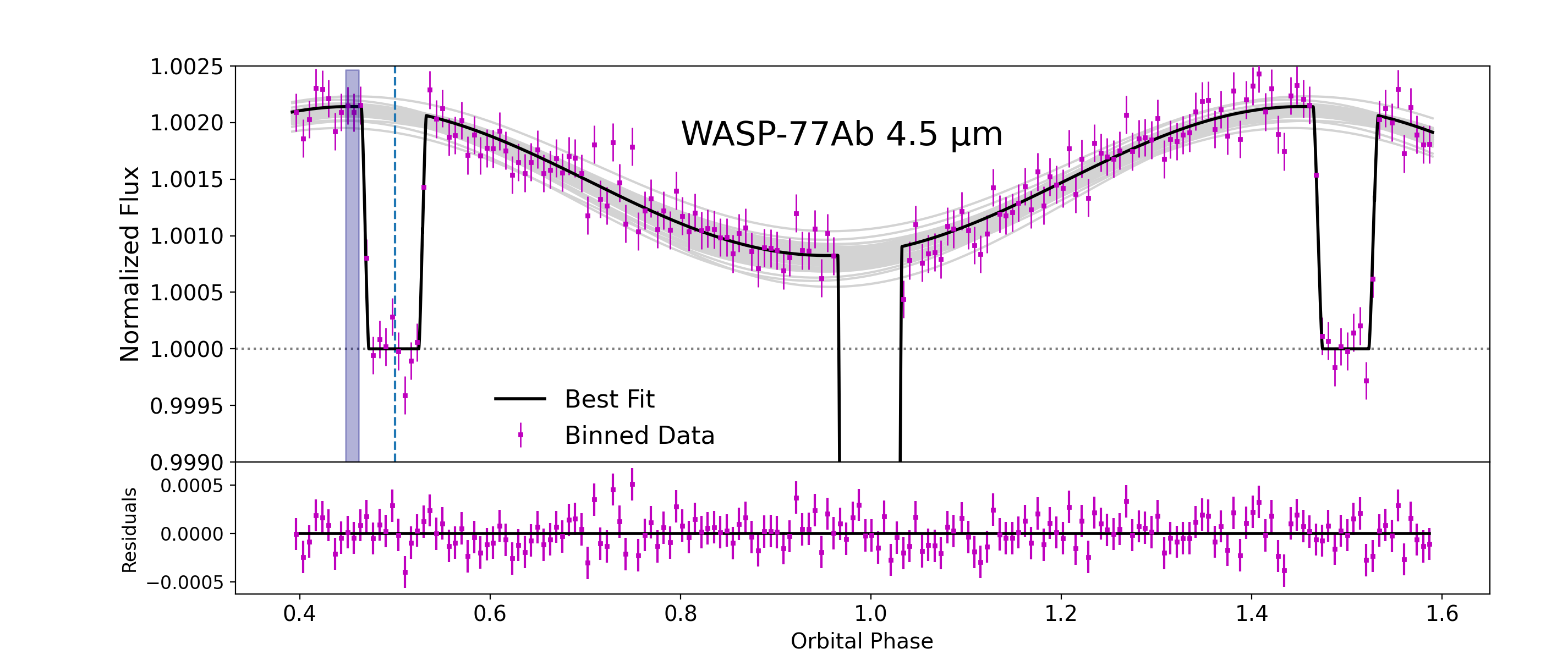}
    \caption{Best fit phase curves for WASP-77Ab at 3.6 $\mu$m (\textbf{top}) and 4.5 $\mu$m (\textbf{bottom}), with resulting residuals. The black curves show the best fit phase, eclipse and transit models. The gray curves are a sampling of 100 MCMC solutions for phase curve parameters to visualize the solution error. The dotted vertical line shows the center of first eclipse, and the vertical blue bar indicates calculated phase offset with error.}
    \label{fig:WASP-77Ab PC}
\end{figure*}

At 4.5 $\mu$m, we found a phase offset of 16.28$\pm$ 2.52$\degree$ with a phase amplitude of 919 $\pm$ 40 ppm. The 4.5 $\mu$m phase offset similarly agrees within 0.6$\sigma$ with the NRS2 estimation of \cite{August2023}. We calculated a corrected eclipse depth of 2947 $\pm$ 144 ppm, resulting in a dayside brightness temperature of 1334 $\pm$ 15 K. On the nightside, we found a brightness temperature of 987$\pm$ 15K. The best fit values are summarized in Table \ref{table:Results}, and plots of the model versus data for both channels are shown in Figure \ref{fig:WASP-77Ab PC}.

\begin{table*}[]
\tabletypesize{\scriptsize}
\caption{Fit Results}
\label{table:Results}
\hspace{-2.5cm}\begin{tabular}{|cc|c|c|c|c|c|c|}
\hline
\multicolumn{2}{|c|}{\multirow{2}{*}{}}                            & \multirow{2}{*}{wa121bo11} & \multirow{2}{*}{wa121bo21} & \multicolumn{2}{c|}{\multirow{2}{*}{wa77abo11}} & \multicolumn{2}{c|}{\multirow{2}{*}{wa77abo21}} \\
\multicolumn{2}{|c|}{}                                             &                            &                            &   \multicolumn{2}{c|}{}                         &    \multicolumn{2}{c|}{}                        \\ \hline \hline
\multicolumn{2}{|c|}{Eclipse Depth {[}ppm{]}}                      &     4077 $\pm$ 59           &     5121 $\pm$ 76                &     \multicolumn{2}{c|}{2507 $\pm$ 59}          &     \multicolumn{2}{c|}{2947 $\pm$ 144}                  \\ \hline
\multicolumn{2}{|c|}{Transit Depth [R$_p$/R$_S$]}                   &   0.12284 $\pm$ 0.00081                      &      0.12402 $\pm$ 0.00042               &      \multicolumn{2}{c|}{0.1110 $\pm$ 0.00029}            &    \multicolumn{2}{c|}{0.1130 $\pm$ 0.00071}                   \\ \hline
\multicolumn{2}{|c|}{Phase Offset [$^{\circ}$]}                               &   -0.78 $\pm$ 1.87                       &   0.42 $\pm$ 1.74                         & \multicolumn{2}{c|}{33.45 $\pm$ 2.79}                      &     \multicolumn{2}{c|}{16.28 $\pm$ 2.52}                      \\ \hline
\multicolumn{2}{|c|}{Amplitude {[}ppm{]}}                          &   1771 $\pm$ 95                           &    2048 $\pm$ 109                   &          \multicolumn{2}{c|}{535 $\pm$ 52}      &       \multicolumn{2}{c|}{919 $\pm$ 40}               \\ \hline
\multicolumn{1}{|c|}{\multirow{4}{*}{F$_p$/F$_S$  {[}ppm{]}}}     & Max   &      4085 $\pm$ 96               &   5141  $\pm$ 107                       &                  \multicolumn{1}{c|}{2697 $\pm$ 60}    & \textit{1907 $\pm$ 42}  &  \multicolumn{1}{c|}{2990 $\pm$ 59} & \textit{2142 $\pm$ 39}                        \\ \cline{2-8} 
\multicolumn{1}{|c|}{}                                     & Min   &      553 $\pm$ 96            &    1045 $\pm$ 107                        &   \multicolumn{1}{c|}{1365 $\pm$ 94}          & \textit{965 $\pm$ 66}     &   \multicolumn{1}{c|}{1152 $\pm$ 55} & \textit{825 $\pm$ 39}                      \\ \cline{2-8} 
\multicolumn{1}{|c|}{}                                     & Day   &   4074 $\pm$ 60              &     5116 $\pm$ 75                       &     \multicolumn{1}{c|}{2510 $\pm$ 61} & \textit{1775 $\pm$ 43}                 &    \multicolumn{1}{c|}{2944 $\pm$ 76} & \textit{2109 $\pm$ 52}                     \\ \cline{2-8} 
\multicolumn{1}{|c|}{}                                     & Night &   554  $\pm$  96                 &         1046 $\pm$ 107                   &      \multicolumn{1}{c|}{1371 $\pm$ 94} & \textit{970 $\pm$ 66}                  &     \multicolumn{1}{c|}{1190 $\pm$ 50} & \textit{853 $\pm$ 35}                \\ \hline
\multicolumn{1}{|c|}{\multirow{4}{*}{Temperature {[}K{]}}} & Max   &    2783 $\pm$ 48               &   2913 $\pm$ 58                         &        \multicolumn{2}{c|}{1939 $\pm$ 26}         &        \multicolumn{2}{c|}{1793 $\pm$ 22}                    \\ \cline{2-8} 
\multicolumn{1}{|c|}{}                                     & Min   &       1259 $\pm$ 67               &   1348 $\pm$ 54                         &           \multicolumn{2}{c|}{1497 $\pm$ 21}            &    \multicolumn{2}{c|}{1221 $\pm$ 22}                   \\ \cline{2-8} 
\multicolumn{1}{|c|}{}                                     & Day   &      2779 $\pm$ 40                 &      2905 $\pm$ 51                 &        \multicolumn{2}{c|}{1876   $\pm$ 23}               &    \multicolumn{2}{c|}{1780 $\pm$ 25}                     \\ \cline{2-8} 
\multicolumn{1}{|c|}{}                                     & Night &   1259   $\pm$    67           &     1349 $\pm$ 54                       &       \multicolumn{2}{c|}{1501 $\pm$ 22}                &      \multicolumn{2}{c|}{1234 $\pm$ 20}              \\ \hline
\end{tabular}
\tablecomments{Eclipse depth, amplitude, R$_p$/R$_S$, F$_p$/F$_S$, and temperatures for WASP-77Ab are adjusted by the dilution factor. Values in italics are prior to dilution correction.}
\end{table*}

\section{Modeling} \label{sec:GCMs}
We performed GCM modeling for WASP-121b and WASP-77Ab using the RM-GCM \citep{Rauscher2010,Rauscher2012}. The dynamical core of the RM-GCM solves the primitive equations of meteorology in pseudo-spectral space, and is coupled to a two-stream radiative transfer routine \citep{Toon1989}. For the models presented here, picket fence radiative transfer was used \citep{Parmentier2015,Malsky2024}. The RM-GCM uses a kinematic MHD approach, which treats the influence of magnetism on the circulation as a Rayleigh drag on the east-west winds (for a global magnetic field aligned with the rotation axis). The timescale for the imposed drag depends on the assumed magnetic field strength and the local electrical resistivity, which can be calculated from the temperature \citep[see details in][]{Rauscher2013,Beltz2022}. The local ionization fractions are calculated from the Saha equation, as per \cite{Menou2012}, resulting in resistivities of $\eta \sim 10^{16}$ cm$^2$ s$^{-1}$ around 1000 K (effectively neutral) to $\eta < 10^{10}$ cm$^2$ s$^{-1}$ for temperatures above 3000 K. In order to feasibly couple these magnetic effects with radiative transfer and clouds, we are necessarily using this simplified approach, which neglects the more complex magnetohydrodynamic effects that can come into play in the hottest regions of the upper atmosphere, where the magnetic Reynolds number can exceed one and the atmosphere can induce its own component of the magnetic field \citep[e.g.,][]{Rogers2017}.
There is also a prescription for radiatively active clouds, which form and dissipate based on temperature conditions as the simulation progresses and contribute to opacities, allowing for radiative feedback \citep{Roman2019,Roman2021,Malsky2024}. Clouds are allowed to be present where local temperature-pressure conditions fall below a species' condensation curve. Maximum cloud thickness in hot Jupiter atmospheres is uncertain and in reality determined by microphysical processes such as gravitational settling and vertical mixing \citep[see, e.g.][]{Powell2024}. For computational efficiency, our model does not treat these processes explicitly, and maximum cloud thickness is prescribed by the user, applied jointly with the thermodynamic constraint \citep[as in][]{Roman2021,Malsky2024}. Four simulations were run with the GCM for each planet comparison to these data: one with no clouds and no magnetic field, one with clouds but no magnetic field, one with no clouds and a 3 G magnetic field, and one with both clouds and a 3 G magnetic field. In the cloudy simulations, eight cloud species were included \citep[those that are not nucleation limited in their formation][]{Malsky2024} and a thickness limit of $\sim 3$ scale heights was placed on each cloud species in each column of atmosphere. All simulations were initialized with a single global-average double-gray profile \citep{Guillot2010} and no winds, and were allowed to spin up for 1000 planet days. Simulations assumed a solar-composition atmosphere and no rainout of TiO or VO. The planets were assumed to be tidally synchronized. All simulations were run with T31 horizontal resolution and 50 logarithmically-spaced pressure levels from 10$^2$ bar to 10$^{-4}$ bar. 

\begin{figure*}
    \centering
    \includegraphics[width=0.8\textwidth]{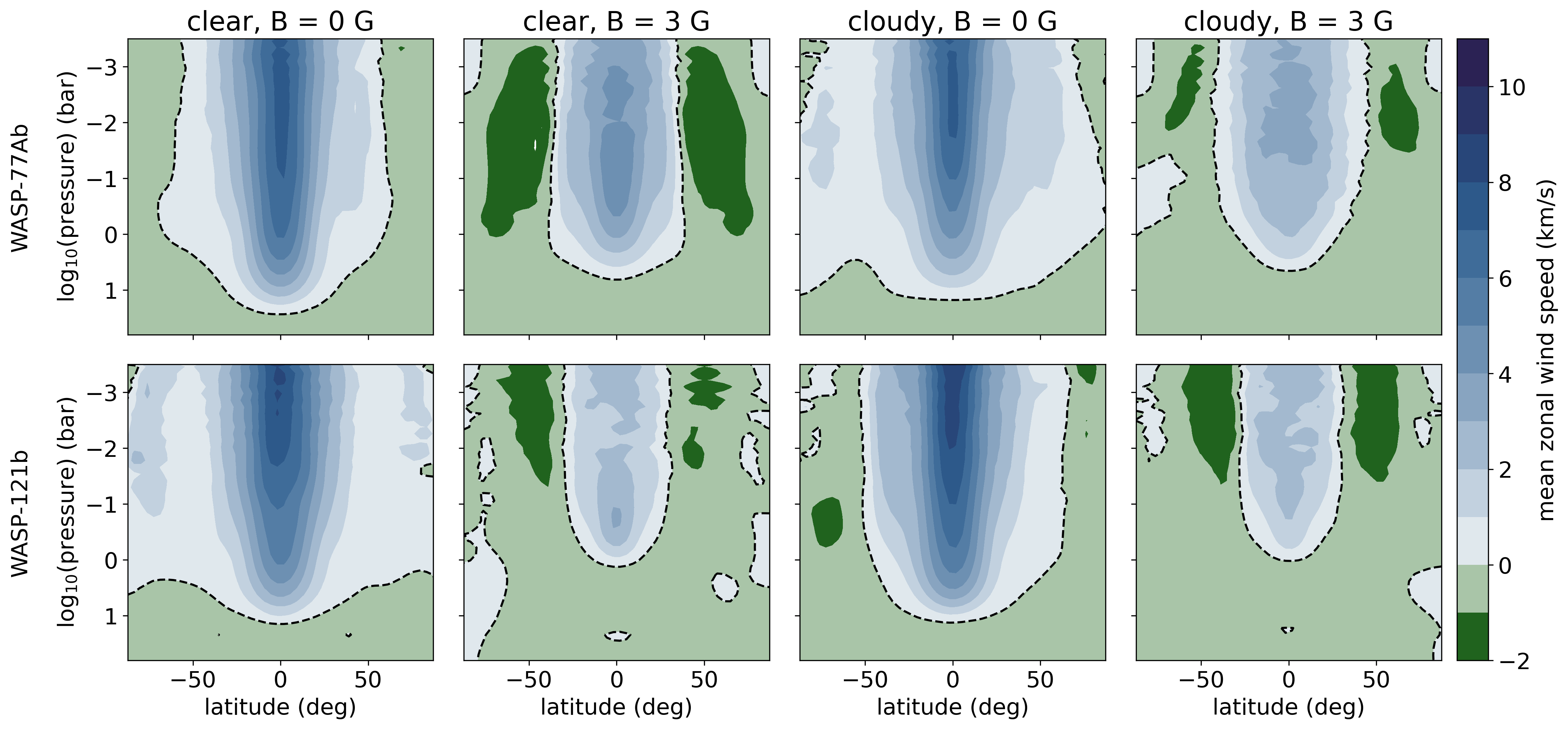}

    \caption{Mean zonal wind speeds (eastward=positive) in each of the GCM models used in this work as a function of latitude and pressure. Different columns represent our different model cases, and the rows correspond to WASP-77Ab and WASP-121b, respectively. Models without magnetic drag develop strong equatorial jets. Including magnetic drag disrupts these jets on the dayside, where the ionization fraction is significant, reducing the strength of the mean zonal wind. For each case we see generally low vertical velocity gradients at 0.1 bar, close to where both Spitzer channels probe.}
    \label{fig:Winds}
\end{figure*}

Figure \ref{fig:GCM} shows the resultant atmospheric structure at the 0.1 bar pressure level, representative of the thermal photosphere for WASP-121b's surface gravity \citep{Fortney2018}. Clouds, when included in the model, are entirely absent on the dayside, but thin clouds above the clear infrared photosphere are present on the nightside. Magnetic effects have a more dramatic effect, as the disruption of the eastward equatorial jet removes the otherwise prominent eastward hotspot shift. This drag also hampers redistribution of heat generally, increasing the day to night contrast. This effect on the wind as both a function of pressure and latitude is shown in Figure \ref{fig:Winds}.

\section{Data-Model Comparison} \label{sec:ModelDataComp}
\subsection{Simulated Phase Curves}
To generate planetary emission spectra from the GCM results, we applied a radiative transfer routine that uses correct line of sight geometries and line-by-line opacities to calculate the emergent flux from each point in the atmosphere at R = 10,000 \citep[as detailed in][]{Kempton2012,Zhang2017,Malsky2021,Savel2022}. Integrating over the disk then produces a planetary emission spectrum for a given phase.

We created simulated phase curves from the emission spectra calculated at each phase by integrating the spectra across the wavelength band of each Spitzer channel, accounting for instrument throughput, and dividing by a simulated stellar flux, similarly converted from a best-match PHOENIX spectrum for each host star \citep{Husser2013}, to produce a normalized result. This led to four model phase curves for each planet, covering all cases with and without clouds and magnetic effects, for each channel. We then compared the model phase curves to our best fit for the observations to find which assumptions best matched the data.  

\begin{figure*}
    \centering
    \includegraphics[width=0.8\textwidth]{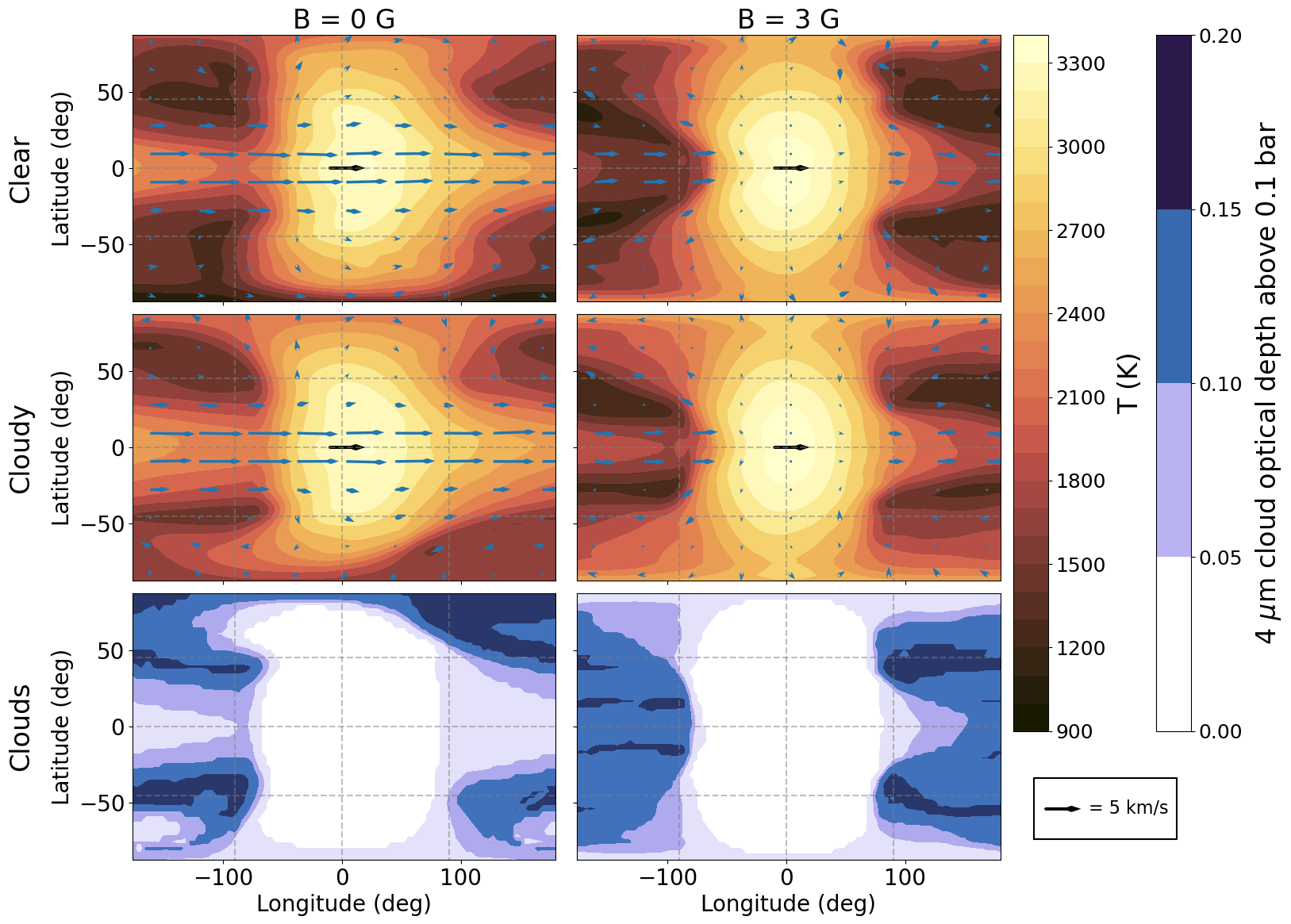}

    \caption{Temperature maps at the 0.1 bar pressure level for each of the WASP-121b GCMs, with wind patterns overlaid. For the two cloudy models, contours of integrated 4 $\mu m$ cloud optical depth above 0.1 bar are displayed in the bottom row. In magnetic models, the eastward equatorial jet is significantly disrupted, reducing the hotspot offset and decreasing the efficiency of heat transport to the nightside, increasing day-night contrast. In cloudy models, clouds are entirely absent from the dayside but present on much of the nightside above 0.1 bar.}
    \label{fig:GCM}
\end{figure*}

\subsubsection{WASP-121b}
\begin{figure*}
    \centering
    \includegraphics[width=0.45\textwidth] { 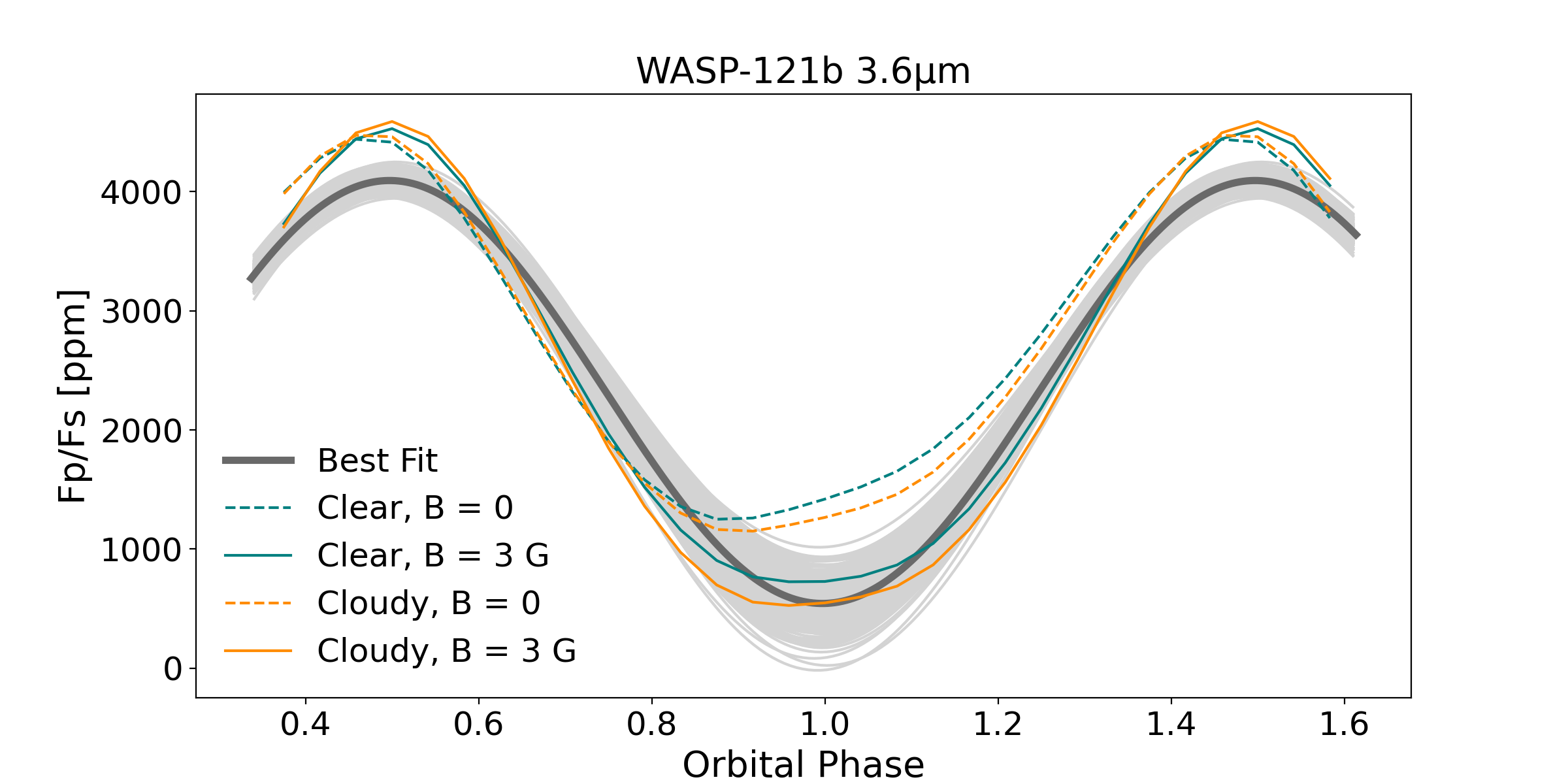}
    \includegraphics[width=0.45\textwidth]{ 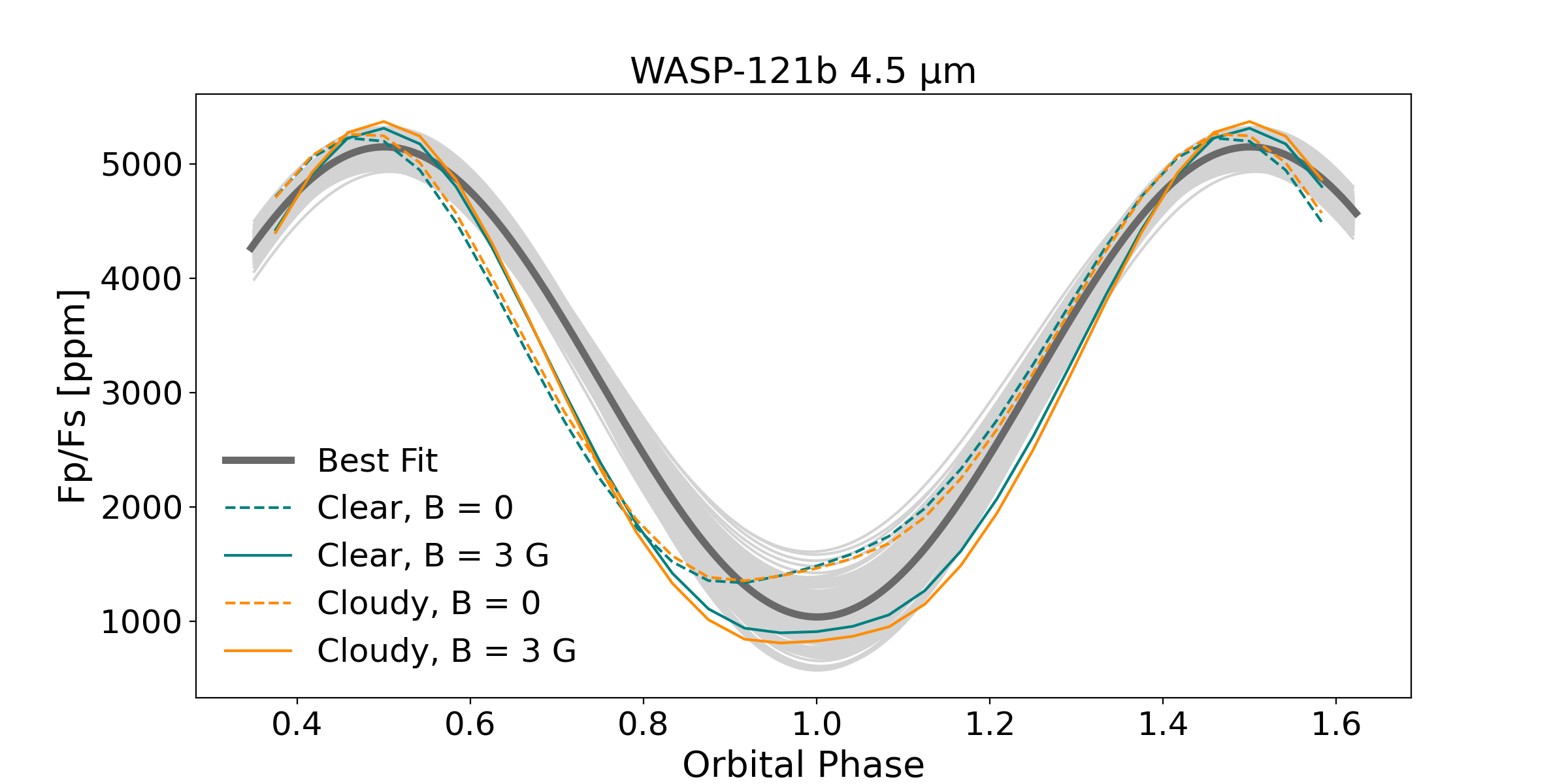}
    \includegraphics[width=0.45\textwidth]{ 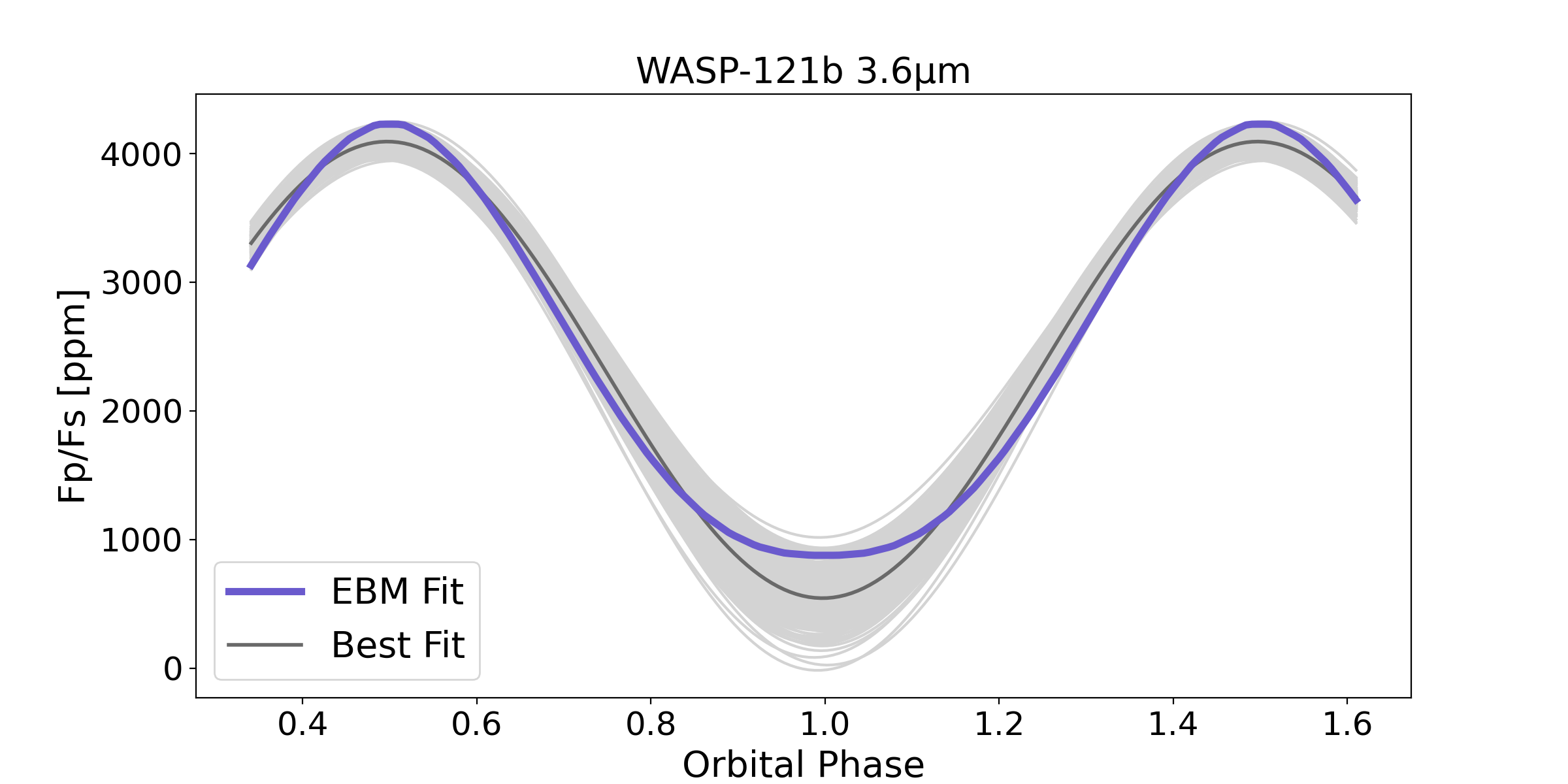}
    \includegraphics[width=0.45\textwidth]{ 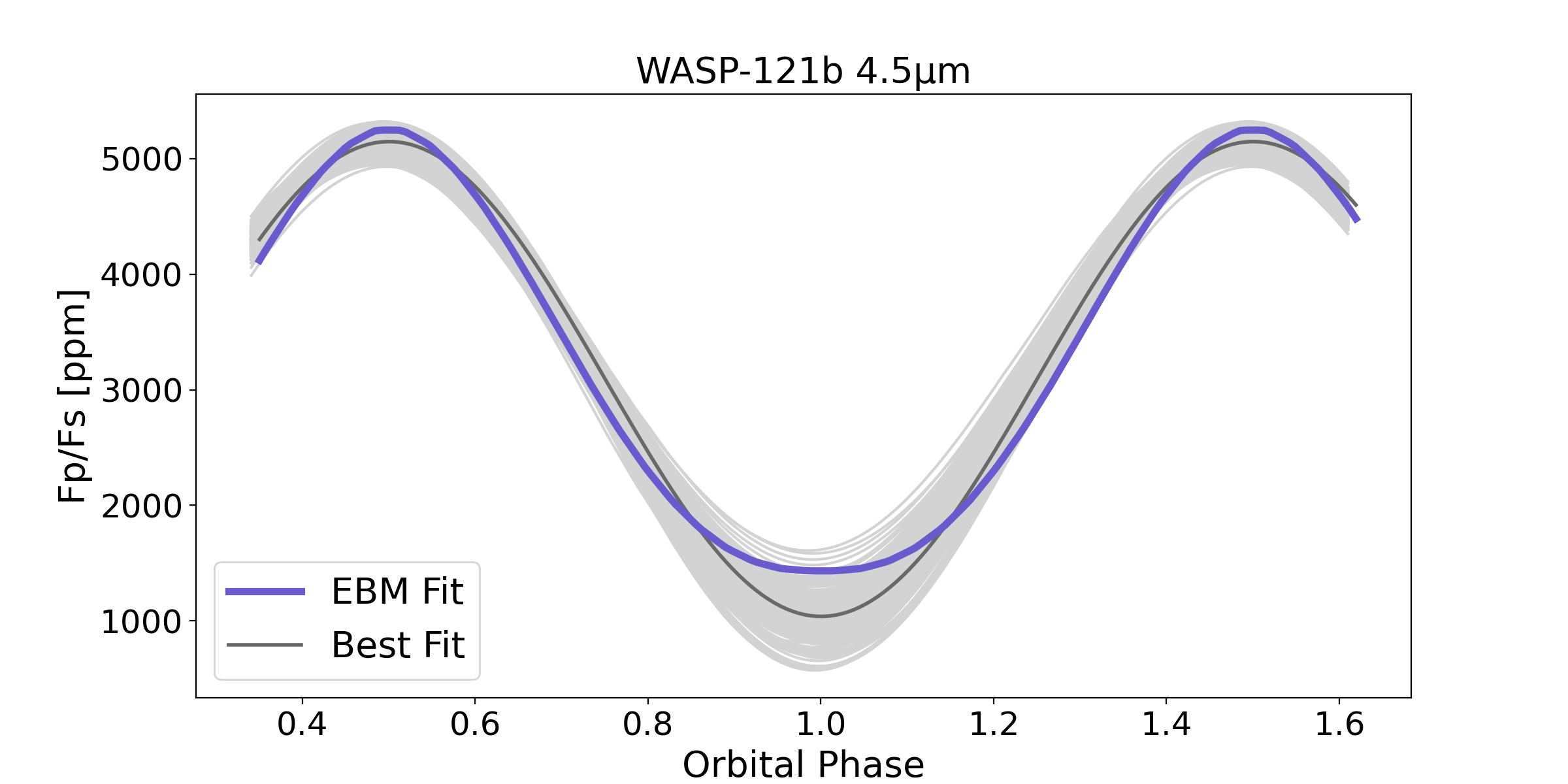}
    \caption{Comparison of the data best fit results at 3.6 and 4.5 µm to modeled phase curves for WASP-121b. The top row compares to RM-GCM results with and without clouds and magnetic effects. The bottom row compares to \texttt{Bell$\_$EBM} fits.}
    \label{fig:Model v Data W121}
\end{figure*}

The inclusion of magnetic effects in the models had the greatest impact on both phase amplitude (increasing it) and phase offset (decreasing it) for both channels. The models with magnetic circulation consequently compared more favorably to the data at both 3.6 µm and 4.5 µm, as visibly evident in Figure \ref{fig:Model v Data W121}. The RM-GCM model phase curves do exhibit distinguishable differences between the clear and cloudy cases, most notably on the nightside where clouds on an ultrahot Jupiter are much more likely to form. In this case, the clouds contribute to a cooler nightside and an unaffected dayside, indicating that we do expect them to form on the nightside of the planet and increase the temperature difference between the two hemispheres while minimally effecting already-damped phase offset. These differences, however, are well within the uncertainty of the Spitzer data and so we are unable to contribute empirical evidence toward the presence of nightside clouds on this planet from these data. 

One identifiable difference between the data and the models is the higher dayside flux for all GCM curves, especially at 3.6 µm. This may largely be due to the fact that the RM-GCM does not include hydrogen dissociation and recombination. \cite{Tan2020} found that the cooling effect of dissociation on ultrahot Jupiters' daysides, and subsequent nightside warming from recombination, serves to reduce the amplitude of phase curves compared to when this effect is ignored. In the equilibrium temperature regimes closest to that of WASP-121b, and when strong drag is included to similar effect as the magnetic circulation of the RM-GCM, they show that the change in flux is more pronounced on the dayside than on the nightside, as the bulk of the recombination occurs at the terminator and weak winds fail to carry the resultant heat to the full nightside. This explains well why the nightside fluxes are in better agreement between the models and data.

When comparing the data to the models, we should keep in mind additional numerical and physical effects that could influence the model predictions.
Three-dimensional modeling of hot Jupiter atmospheres is challenging, as it stretches our classic understanding of atmospheric dynamics to a new, extreme regime. The dynamics in any atmosphere will be influenced by the strength of forcing and dissipation, but for the intense irradiation and high wind speeds characteristic of hot Jupiter atmospheres, it must be especially recognized that the wind patterns and speeds may be incorrect if the numerical dissipation is too strong \citep{Heng2011,Hammond2022,Christie2024}, especially as the correct amount of dissipation to use will depend on the forcing conditions and spatial resolution used \citep{Thrastarson2011}. Using too coarse of a spatial resolution could result in incorrect flow \citep{Skinner2021}, although for the large dynamical scales within the hot Jupiter regime, most models may be sufficiently resolved \citep{Menou2020}. Then there are additional basic concerns about whether the fluid dynamics approximations and boundary conditions are appropriate \citep[e.g.,][]{Mayne2014}. Given these concerns, it is notable that there exists fairly good qualitative agreement between hot Jupiter GCMs from different groups, using different approaches and assumptions, and the growing set of atmospheric observations, including phase curves \citep[see an explicit discussion of this in the review by][]{Showman2020}. The quantitative mismatch between models and GCMs is often attributed to uncertainties in intrinsic physical parameters (such as atmospheric metallicity) and/or additional physics that was not included in the models. We are testing two of those additions here, clouds and the influence of magnetism, but each brings with it a whole new set of numerical and physical uncertainties \citep[also reviewed in][]{Showman2020}. Extensive testing of physical and numerical uncertainties in the GCMs of these two planets is beyond the scope of this paper; instead we present these models for use in the initial interpretation of the phase curves. The use of EBMs, below, gives us an alternate method for interpretation, relying on fewer complexities at the cost of being less physically motivated.

\subsubsection{WASP-77Ab}
For WASP-77Ab, no GCM run came close enough to the data to warrant detailed comparison between different model cases. Figure \ref{fig:Model v Data W77} shows the excessive phase amplitude and peak thermal fluxes when compared against both Spitzer phase curves. Phase amplitudes from the RM-GCM runs averaged 120\% higher than our fit to the data at 3.6 µm, and 53\% higher at 4.5 µm. (the closest model disagrees by 9$\sigma$ in amplitude with the data fit for both channels). We attribute this to the low metallicity of WASP-77Ab, as identified by \cite{Line2021}, \cite{August2023}, and \cite{Smith2024}. The RM-GCM is currently constrained to solar and supersolar metallicity modeling, which for this planet results in a temperature inversion in the modeled temperature-pressure profile. This leads to unreasonably high fluxes in the 3-5 µm region when compared to eclipse depth measurements from Spitzer in this work and by \cite{Mansfield2022}, and to the best-fit thermal emission spectrum for JWST NIRSpec data by \cite{August2023}. The changes necessary to enable low-metallicity modeling are beyond the scope of this work.

\subsubsection{Energy Balance Model}
In order to employ a consistent model which would allow both an analysis of the WASP-77Ab phase curve and a direct comparison between the two planets observed in this work, we ran an Energy Balance Model (EBM) at 3.6µm and 4.5 µm for each planet, wholly independent of the GCMs. The EBM provides a more simplified approach to characterizing planetary influences on the amplitude and offset of phase curves.

\paragraph{EBM Description}

For our energy balance model, we utilized \texttt{Bell$\_$EBM} \citep{Bell2018}, which takes as inputs orbital, stellar, and intrinsic planetary properties and produces phase curves by calculating planet-to-star flux ratios at different orbital phases assuming a planet in global radiative equilibrium. In our analysis, we set the parameters in Tables \ref{table:planet params} and \ref{table:LimbDarkening} as fixed values. 

\begin{figure*}
    \centering
    \includegraphics[width=0.45\textwidth] { 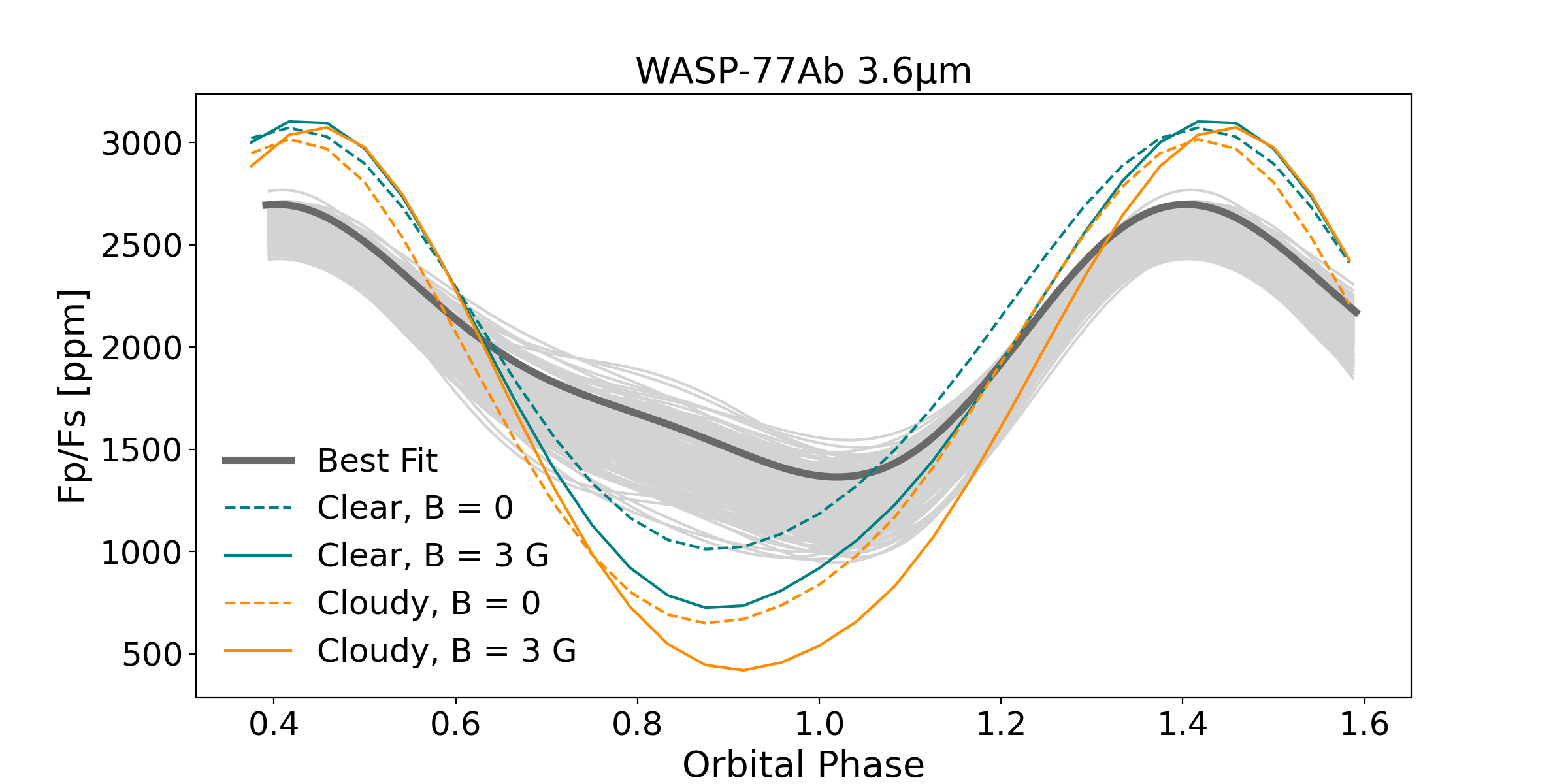}
    \includegraphics[width=0.45\textwidth]{ 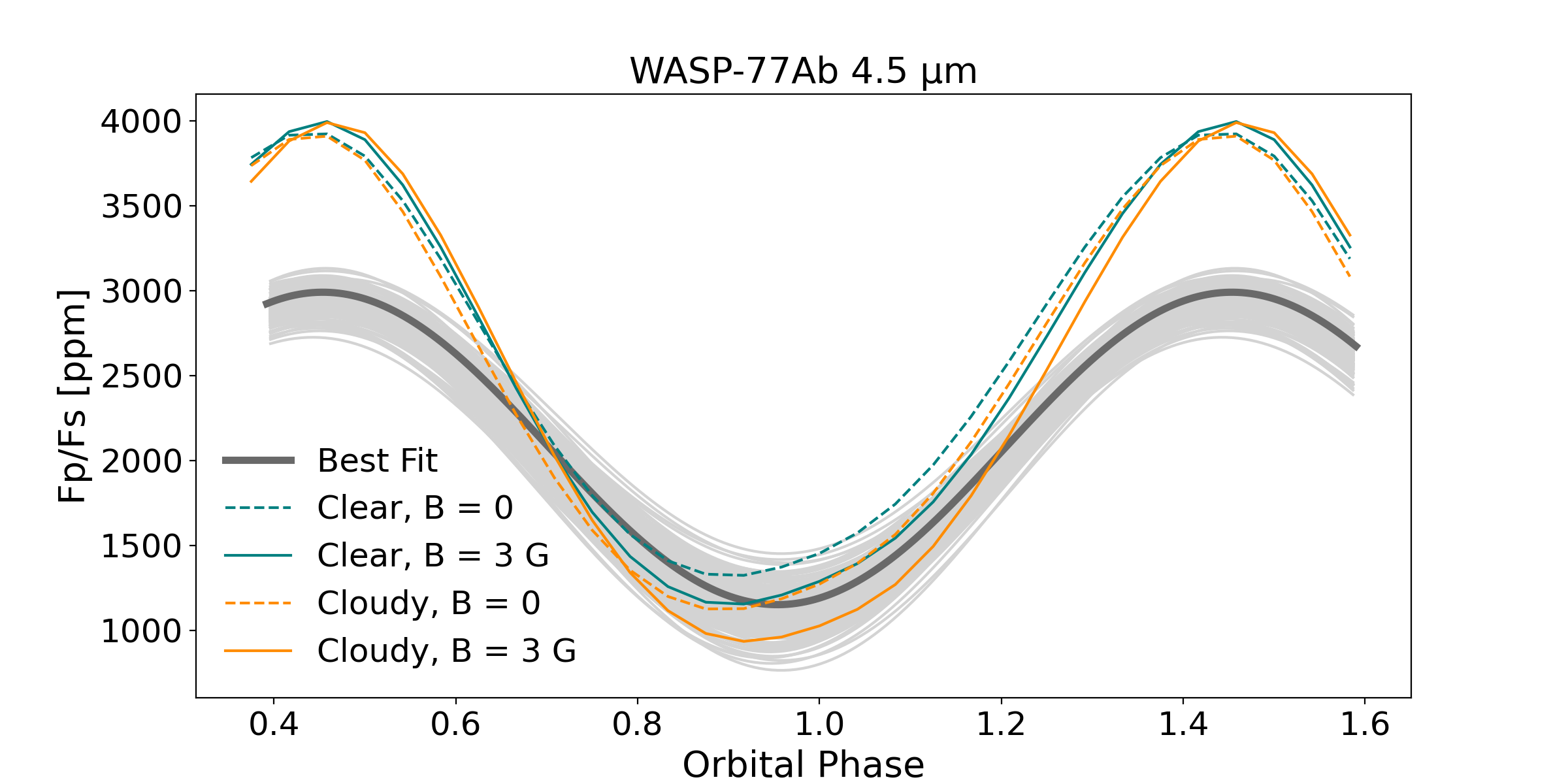}
    \includegraphics[width=0.45\textwidth]{ 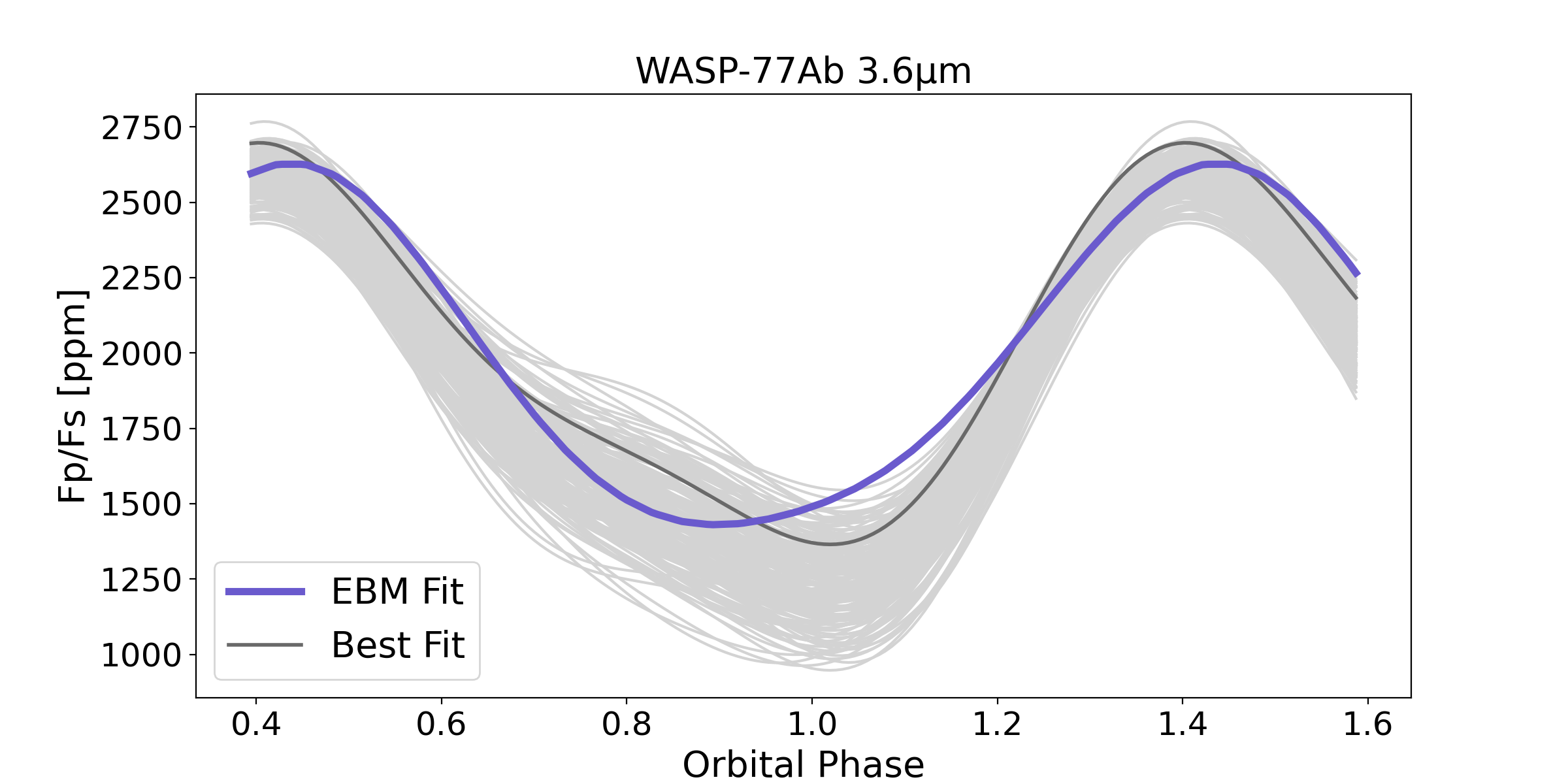}
    \includegraphics[width=0.45\textwidth]{ 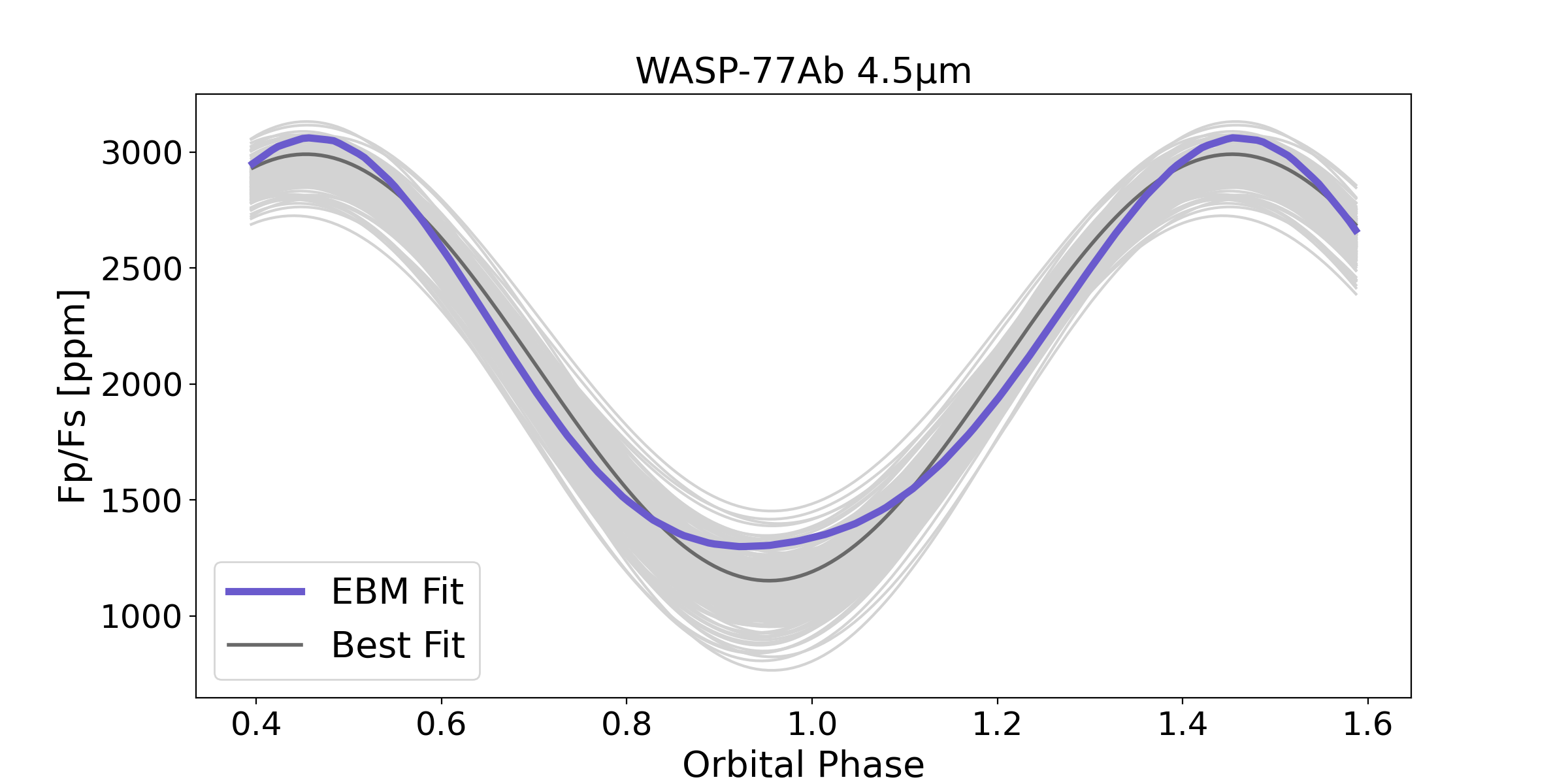}
    \caption{Comparison of our best fit to the data with model results, similarly to Figure \ref{fig:Model v Data W121}, for WASP-77Ab. Phase amplitudes and dayside thermal fluxes from RM-GCM results are excessive compared to the data. The simplified \texttt{Bell$\_$EBM} is able to reproduce measured fluxes by assuming an atypically high Bond albedo.}
    \label{fig:Model v Data W77}
\end{figure*}

We then performed an MCMC fitting scheme to find which atmospheric properties most closely reproduce our phase curve fits for the observational data, presented in Figures \ref{fig:WASP-121b PC} and \ref{fig:WASP-77Ab PC}.  To accomplish this, we fit for four atmospheric parameters: Bond albedo ($A_B$), zonal wind speed (converted from angular frequency of the atmosphere rotating as a solid body around the planet), internal flux, and mixed layer depth (in pressure, as $P_0$). Each of these parameters contributes to the planet flux calculated at each time-step along the planets orbit, which is based on the change in temperature at each latitude/longitude pixel of the planet due to both radiation and advection. Albedo and internal flux drive the magnitude of radiative heating, while the radiative timescale is adjusted by the mixed layer depth, which scales the atmosphere's heat capacity. The dynamical timescale is set by the wind speed. The resulting best-fit values are outlined in Table \ref{table:EBM} and the lightcurves in comparison to the data fits are plotted in Figures \ref{fig:Model v Data W121} and \ref{fig:Model v Data W77}. Calculating fluxes at 3.6 µm and 4.5 µm, versus bolometrically, does introduce further approximations, such as a globally uniform photosphere at each wavelength, but it allows us to evaluate the scales of each parameter necessary for the model to best recreate the data. 

\paragraph{EBM Results}

The EBM is able to reproduce the observed phase curves to within our uncertainties for both planets. The WASP-121b fits for both photometric bands are consistent with suppressed wind speeds of under 100 m/s (see Table \ref{table:EBM} as necessary to explain the near-zero phase offset in the observation. The nightside fluxes, however, are overestimated in comparison to the data fits. For WASP-77Ab, we see unsurprising (1-2 km/s) wind speeds in order to achieve the moderate phase offsets in the Spitzer data. The EBM requires an unexpectedly high Bond albedo, though, in order to fit the low fluxes and phase amplitudes. Moderate Bond albedos, typically up to 0.4, have precedent in hot Jupiters, which has been explained by the presence of clouds that significantly reflect near-infrared light \citep{Mallonn2019}. The albedo fits for WASP-77Ab, 0.601 at 3.6 µm and 0.582 at 4.5 µm, are even higher, and although this may be in part attributable to such clouds, the albedo term may be acting as a catch-all for any effects not modeled which have a greater effect on amplitude than on offset. Shorter wavelength observations and GCM analyses with more varied cloud types may help constrain the contribution from cloud cover.

\begin{table*}[]
\caption{EBM Best Fit Values}
\label{table:EBM}
\centering
\begin{tabular}{|c|cc|cc|c|}
\hline
\multicolumn{1}{|c|}{\multirow{2}{*}{Fitted Parameter}} & \multicolumn{2}{c|}{WASP-121b}    & \multicolumn{2}{c|}{WASP-77Ab}        & \multirow{2}{*}{Priors} \\ \cline{2-5}
\multicolumn{1}{|c|}{}                                  & \multicolumn{1}{c|}{3.6 µm} & 4.5 µm &  \multicolumn{1}{c|}{3.6 µm} & 4.5 µm  &
\\ \hline \hline
Bond albedo ($A_b$)                                       & \multicolumn{1}{c|}{ 0.228 $^{+.011}_{-.020 }$}      &  0.002 $^{+.003}_{-.001}$       &   \multicolumn{1}{c|}{0.608 $^{+.021}_{-.019 }$}   &  0.582 $^{+.011}_{-.012}$ &      [0, 1]              \\ \hline
Wind Velocity (m/s)                             & \multicolumn{1}{c|}{43.5 $^{+148}_{-32}$}       &  -14.9 $^{+89}_{-219}$      &  \multicolumn{1}{c|}{2210 $^{+148}_{-32}$}      &  1168 $^{+1922}_{-569}$ &           [-5000, 5000]               \\ \hline
Internal Flux ($W/m^2$)                                         & \multicolumn{1}{c|}{2.63 $^{+.15}_{-.36}$ x $10^5$ }      &   3.85 $\pm$ .12 x $10^5$     &   \multicolumn{1}{c|}{2.49 $\pm$ .11 x $10^5$}      & 1.37 $^{+.06}_{-.07}$x $10^5$   &        [0, 5 x $10^5$]                 \\ \hline
Layer Depth ($P_0$) (bar)                                                     & \multicolumn{1}{c|}{ 0.115 $^{+.248}_{-.101}$}       &   0.017 $^{+.069}_{-.015}$     &   \multicolumn{1}{c|}{0.427 $^{+.329}_{-.183}$}    &  0.362 $^{+.404}_{-.225}$  &      [0, 1]              \\ \hline
\end{tabular}
\caption{Key model parameter values required for the energy balance model to recreate the observed phase curves. Wind velocity and layer depth drive the fit to offset, while bond albedo accounts for additional influences on amplitude. Because this is a simple model, the high albedo fits for WASP-77Ab are likely not reflective of actual planet values, but indicate additional unmodeled effects are influencing the low day-to-night temperature difference.}
\end{table*} 

\subsubsection{GCM-EBM Comparison}
Our RM-GCM results for WASP-121b suggest that magnetic drag reduces both the phase offset and raises the phase amplitude primarily through reducing the nightside flux. As evident in Figure \ref{fig:GCM}, this is caused by the slowing of the zonal equatorial jet and increased meridional flow over the poles, which is less efficient at heat transfer. while on the nightside, where the resistivity is more than 3 orders of magnitude higher, winds remain largely undisturbed by the inclusion of a magnetic field. The \texttt{Bell$\_$EBM} considers only a single value for wind that is applied globally, so reduced dayside zonal winds couples with commensurately reduced nightside winds. This difference, in addition to not including meridional flow, may explain the excess nightside flux present in the EBM results, as well as the low wind speeds in Table \ref{table:EBM} compared to global zonally-averaged values of 400-500 m/s from the GCM results despite near-zero dayside values when including magnetic fields, visualized in Figure \ref{fig:GCM}. Additionally, Figure \ref{fig:Model v Data W121} shows that including clouds will lower the nightside flux while minimally influencing the dayside, where clouds are unlikely to form at its elevated temperatures. Another possibility for the discrepancy is the presence of dayside hydrogen dissociation and nightside recombination, an effect which is included in the EBM runs for this work. Assuming that this process is not occurring would lower the nightside flux, as the recombination process in cooler regions is exothermic; however, because the equilibrium temperature is well within the region where \cite{Tan2020} show this phenomenon is expected to exist and influence phase curve observables, we assess it is likely that this effect is present in WASP-121b.

The EBM results for WASP-77Ab suggest, in addition to a high albedo, the presence of substantial zonal winds as a means to both reduce the dayside-to-nightside flux contrast and to recreate the moderate-to-high phase offsets identified in the data. Though derived from different assumptions, both the EBM (1-2 km/s), and the GCM (1.2 km/s with a magnetic field and 2.5 km/s without at 0.1 bar), suggest roughly similar global wind speeds to match the measured phase offsets.

\subsection{Updates to Population Trends}
\begin{figure}
    \centering
    \includegraphics[width=0.5\textwidth]{ 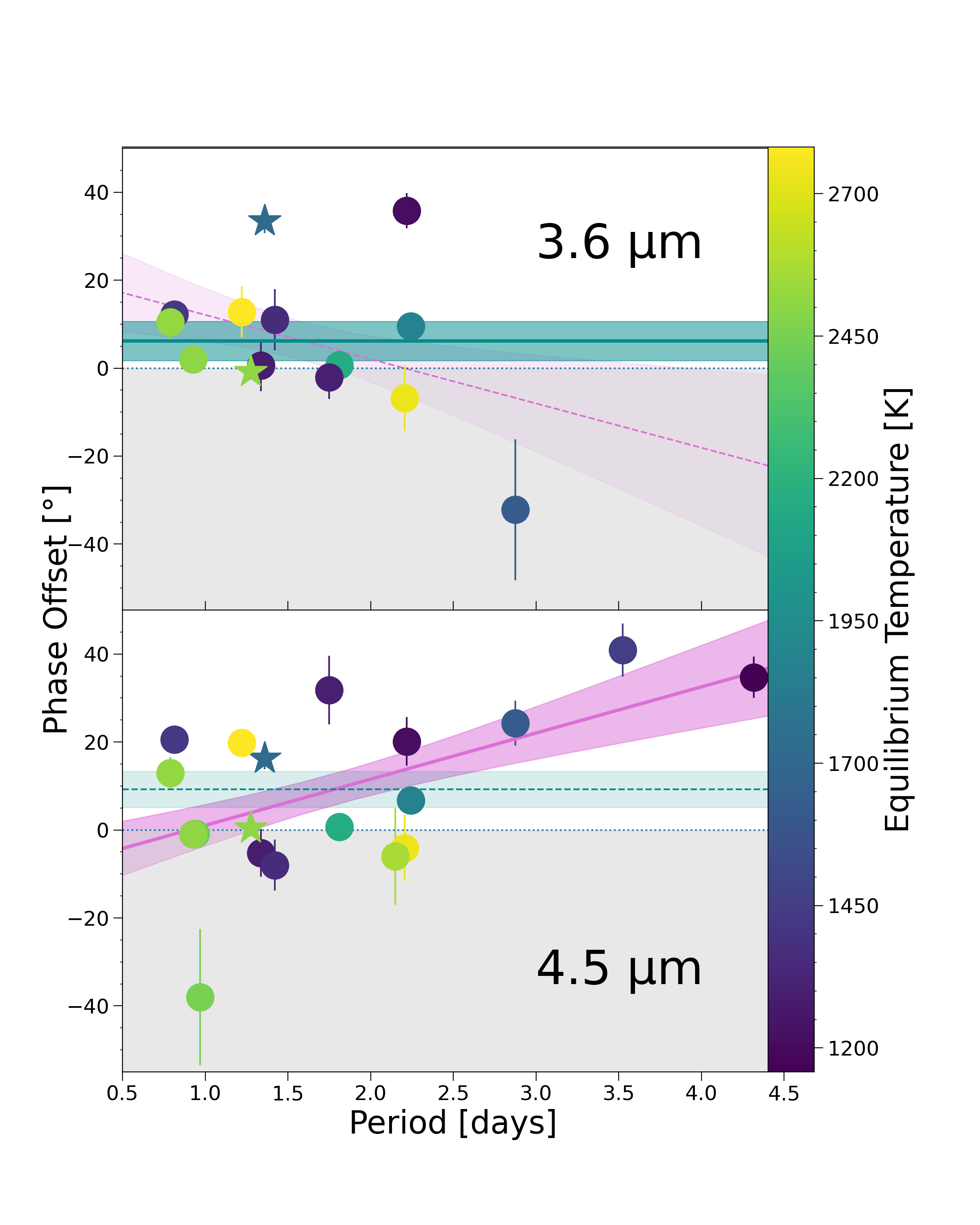}

    \caption{Plot of period versus phase offset for 3.6 $\mu$m (top) and 4.5 $\mu$m (bottom) Spitzer data. The purple lines show calculated best fit trends (with error), and the cyan lines show constant value (no trend) fits. The solid-lined, darker shading highlight the fit preferred by BIC. The stars represent WASP-121b and WASP-77Ab. Each planet is colored by equilibrium temperature.}
    
    \label{fig:Period vs Offset}
\end{figure}

\cite{May2022} tested for a trend in phase offset versus orbital period for a population of 18 hot Jupiters, seven of which were analyzed by a uniform process. Here, we reassess the trend with the addition of WASP-121b and WASP-77Ab, analyzed in as close to uniform as possible a method as the stated seven planets and using the same published values for all other planets as \cite{May2022}, excluding WASP-140b, which was not used for trend analysis in the previous paper due to its eccentric orbit, and WASP-12b due to reported variability in its phase offset \citep{Bell2019}. Comparing a linear fit of phase offset versus period to a constant value, a positive trend is still preferred ($\Delta$BIC = -460.3) although less strongly and with a shallower slope than without the addition of WASP-121b and WASP-77Ab ($\Delta$BIC = -696.9 with neither planet included). Figure \ref{fig:Period vs Offset} shows the period versus phase offset relation for the population with the best fit trend overlayed.

We also explored the phase offset versus period trend for the population at 3.6 $\mu$m using all published results for this Spitzer channel. Here, the best fit line showed a \textit{negative} trend, although it was not preferred to a flat line ($\Delta$BIC = +175.1), suggesting no relation between phase offset and period from 3.6 $\mu$m observations. The the lack of 3.6 µm data on longer-period planets did, however, limit the range of orbital periods considered when compared to the 4.5 $\mu$m data.

\section{Conclusions} \label{sec:Conclusions}
From Spitzer phase curve observations of WASP-121b and WASP-77Ab at both 3.6 µm and 4.5 µm, we fit sinusoidal model phase function, as well as model transits and eclipses, to derive informative planetary observables, namely phase amplitude and offset, and dayside and nightside brightness temperatures, shown in Table \ref{table:Results}. During the data analysis process, we investigated the root cause for Spitzer calibration data precluding the creation of a fixed BLISS map for \texttt{POET} at 3.6 µm, and found evidence of variability in the calibration star. With the Spitzer mission no longer active, and no other existing calibration data at 3.6 µm sufficient to build a fixed BLISS map, we propose continued free BLISS mapping for future \texttt{POET} reductions of 3.6 $\micron$ data.

Following the data reduction, we compared the resulting fits to GCM and EBM modeled phase curves, as well as reassessing data-driven population trends, with the following results:

\begin{itemize}
    \item The low phase offset and large phase curve amplitude of WASP-121b at both 3.6 µm and 4.5 µm together mandate that this planet has poor day-to-night heat redistribution. Since this planet is of a type known to host thick atmospheres, the weak redistribution is consistent with GCM results that have some form of drag reducing day-night heat advection. We recreate this effect in our models through the influence of magnetism shaping the circulation. While our simulated phase curves can additionally distinguish between the presence and absence of clouds, particularly on the planet's nightside, the differences are well within the uncertainty of the Spitzer observations.
    \item WASP-77Ab has an unexpectedly low overall thermal flux and phase curve amplitude, which are difficult to account for in both 3D GCMs and 1D energy balance models. EBM results suggest zonal wind speeds of a few km/s as well as a high Bond albedo, which may be related to cloud cover or to the model's simplified treatment of the factors affecting the planet's incident and emitted light.
    \item An observational trend of increasing phase offset with planet period at 4.5 µm persists, but the data from this work weakens the extant trend. Inclusion of 3.6 µm data further obfuscates the trend, although generally noisier data and a smaller spread in observed periods may contribute.
    \item Phase curve observables measured in this work, including phase offset, amplitude, and secondary eclipse depth, agree well with present JWST results for both planets. This suggests that Spitzer phase curve results are robust, although the reduction method of Spitzer data remains important. Further comparisons are necessary to determine to what extent the Spitzer reduction method influences agreement with JWST phase observations.
\end{itemize}

Future analysis of the dynamics of WASP-77Ab would benefit from the inclusion of subsolar metallicity and from a robust treatment of clouds, both of which may significantly impact heat transport in the planet's atmosphere. 
\pagebreak
\software{\\ Astropy \citep{astropy,astropy2},
\\ batman \citep{Kreidberg2015},
\\ ExoCTK \citep{exoctk},
\\ IPython \citep{ipython},
\\ Matplotlib \citep{matplotlib},
\\ NumPy \citep{numpy, numpynew},
\\ SciPy \citep{scipy},
}

\acknowledgments
This research has made use of the NASA Exoplanet Archive, which is operated by the California Institute of Technology, under contract with the National Aeronautics and Space Administration under the Exoplanet Exploration Program. This paper makes use of data from the first public release of the WASP data as provided by the WASP consortium and services at the NASA Exoplanet Archive. 

This work was supported by NASA XRP grant 80NSSC22K0313. We thank Thaddeus Komacek and Hayley Beltz for useful conversations about the EBM results.

\bibliography{main}{}
\bibliographystyle{aasjournal}
\end{document}